\documentclass{emulateapj} 
\usepackage{amsmath}
\usepackage{amssymb}
\usepackage{color}
\usepackage{natbib}

\shorttitle{Metal Accretion onto halo main-sequence stars} \shortauthors{Hattori, Yoshii, Beers, Carollo, \& Lee}

\begin{document}

\title{Possible Evidence for Metal Accretion onto the Surfaces of Metal-Poor Main-Sequence Stars} 

\author{
Kohei Hattori\altaffilmark{1}, 
Yuzuru Yoshii\altaffilmark{1}, 
Timothy C. Beers\altaffilmark{2,3},  
Daniela Carollo\altaffilmark{4,5}, and 
Young Sun Lee\altaffilmark{6}}

\altaffiltext{1}{Institute of Astronomy, School of Science, University of Tokyo, 2-21-1, Osawa, Mitaka, Tokyo 181-0015, Japan}
\altaffiltext{2}{National Optical Astronomy Observatories, Tucson, AZ 85719, USA}
\altaffiltext{3}{JINA: Joint Institute for Nuclear Astrophysics}
\altaffiltext{4}{Macquarie University - Dept. Physics \& Astronomy, Sydney, 2109 NSW, Australia}
\altaffiltext{5}{INAF - Osservatorio Astronomico di Torino, 10025 Pino Torinese, Torino - Italy}
\altaffiltext{6}{Department of Astronomy, New Mexico State University, Las Cruces, NM 88003, USA}
\email{khattori@ioa.s.u-tokyo.ac.jp}

\begin{abstract}

The entire evolution of the Milky Way, including its mass-assembly and
star-formation history, is imprinted onto the chemo-dynamical
distribution function of its member stars, $f(x, v, {\rm [X/H]})$, in
the multi-dimensional phase space spanned by position, velocity, and
elemental abundance ratios. In particular, the chemo-dynamical
distribution functions for low-mass stars (e.g., G- or K-type dwarfs)
are precious tracers of the earliest stages of the Milky Way's
formation, since their main-sequence lifetimes approach or exceed the
age of the universe. A basic tenet of essentially all previous analyses
is that the stellar metallicity, usually parametrized as [Fe/H], is
conserved over time for main-sequence stars (at least those that have
not been polluted due to mass transfer from binary companions). If this
holds true, any correlations between metallicity and kinematics for
long-lived main-sequence stars of different masses, effective
temperatures, or spectral types must strictly be the same, since they
reflect the same mass-assembly and star-formation histories. By
analyzing a sample of nearby metal-poor halo and thick-disk stars on the
main sequence, taken from Data Release 8 of the Sloan Digital Sky
Survey, we find that the median metallicity of G-type dwarfs is
systematically higher (by about 0.2 dex) than that of K-type dwarfs
having the same median rotational velocity about the Galactic center. If
it can be confirmed, this finding may invalidate the long-accepted
assumption that the atmospheric metallicities of long-lived stars are
conserved over time. 

\end{abstract}


\section{Introduction}\label{introduction}

Numerical calculations of the nuclear processes that take place in
stellar interiors have established that the evolutionary track of a
given star is determined primarily by two parameters -- its mass and
chemical composition (other parameters, such as stellar rotation and
magnetic fields may play roles as well, but they are expected to be
smaller). While main-sequence stars are believed to retain the initial
values of their mass and composition from birth through the completion
of core H burning, it has been conjectured previously that this may not
necessarily be the case, depending on their surrounding environment. For
example, \cite{Bondi1952} demonstrated, from theoretical considerations,
that a star can accrete gas onto its surface at a rate proportional to
$v_{\rm rel}^{-3}$, where $v_{\rm rel}$ is the relative velocity between
the star and the gas with which it collides. In particular, if initially
metal-poor stars accrete metal-rich gas, this would lead to an
enhancement of the metallicity in their atmospheres, and thus confound a
straightforward interpretation of the chemical evolution of the Galaxy. 

\cite{Yoshii1981} was the first to speculate that Bondi mass accretion 
onto halo stars may drastically alter their surface metal abundances in
the course of formation of the Galactic halo, proposing that more
metal-poor halo stars would suffer greater surface metal enhancements
than metal-rich stars, due to their shallower surface convective
envelopes. He argued that, early in the formation history of the Milky
Way, before the emergence of the disk, gaseous material is primarily
distributed throughout the halo. During this era, high-density gas
clouds move more or less randomly throughout the halo. If halo stars
formed from these clouds, it is possible for them to collide at small
relative velocity with chemically processed dense gas from previous
star-formation episodes in these same clouds, so that mass accretion as
well as surface metal enhancement is likely to occur. After the halo gas
has contracted to form the disk stellar populations, mass accretion onto
halo stars becomes negligible, for two reasons. First, halo stars
collide with dense gas only occasionally, when they pass through the
disk. Secondly, such collisions occur near the pericenter of the stellar
orbits, hence the relative velocity between halo stars and the disk gas
is too large to enable efficient mass accretion. 

According to the currently favored hierarchical galaxy formation
paradigm in a $\Lambda$-CDM universe, the Galaxy is thought to form
through mergers of sub-galactic systems for which the internal velocity
dispersion is much smaller than that of the ensemble of these systems
that make up the entire halo (see, e.g., \citealt{Tissera2013}, and
references therein). Because the relative velocity between the stars and
the gas within these systems is on the order of their internal velocity
dispersions, this scenario implicitly involves conditions that are
favorable for mass accretion, as suggested years ago by \cite{Yoshii1981}. 

Several previous authors have attempted to evaluate the rate of mass
accretion in a cosmological context, and reached similar conclusions,
that the surface metal abundances of low-mass, initially metal-poor (or
even near zero-metallicity stars) could indeed be enhanced significantly (e.g.,
\citealt{Shigeyama2003, Komiya2010}). Other authors have argued 
against the notion that significant mass accretion onto halo stars could
have occurred. For example, \cite{Frebel2009} calculated the orbits of
individual halo stars in order to evaluate the amount of mass that might
have been accreted as they penetrate the disk (several times over the
age of the Galaxy), and concluded that their surface metal abundances are
hardly enhanced.  However, in this work, the amount of mass accreted
by individual stars within their natal sub-galactic systems, which may be
the dominant source, is not taken into account.

In order to break this stalemate, we propose an observational test to
decide whether or not the accretion hypothesis may indeed be supported.
Suppose that metal-poor halo G- and K-type dwarfs, with main-sequence
lifetimes close to or exceeding the age of the universe, experience
no mass accretion and associated surface metallicity
enhancement, and that their kinematical properties are independent of
spectral type, because they share the same star-formation and
mass-assembly histories. Then, any observed correlation between
metallicity and kinematics for G-type dwarfs must be identical to that
observed for K-type dwarfs. However, if mass accretion and metallicity
enhancement did occur, the situation is expected to be quite different.
According to theoretical studies of the internal structures of stars,
the mass of the surface convective envelope for dwarfs drastically
decreases with increasing effective temperature, $T_{\rm eff}$. This
implies that the affect of surface metal enhancement on G-type dwarfs is
expected to be much larger than for K-type dwarfs (due to the lack of dilution
in their surface layers), even if they had the same initial metal abundance
and the same amount of mass accreted. Thus, we can accept or reject the
accretion hypothesis by examining whether a spectral-type dependent
shift in metallicity for stars with otherwise identical kinematics
exists or not. 

Note that this test is possible only statistically, based on a large,
kinematically unbiased sample of metal-poor main-sequence dwarfs. In
this paper, we employ spectroscopic observations of G- and K-type dwarfs
in a relatively local region of the halo, taken from Data Release 8 of
the Sloan Digital Sky Survey (SDSS DR8, \citealt{Aihara2011}), to derive
the relation between the median rotational velocity, $\langle V_\phi \rangle_{\rm med}$, and the
surface metal abundance, [Fe/H], for stars of different spectral types.
If the above mentioned shift in metallicity with respect to kinematics
is confirmed, it would be the first observational evidence of the
importance of mass accretion onto halo stars. Given the implications of
this, and the clear influence on the interpretation of future
observations, it is crucial to seek confirmation (or refutation) based
on additional studies. 

This paper is organized as follows. 
We describe our sample selection in Section 2, 
and present our analysis and results in Section 3. 
The plausibility of our results is discussed in Section 4. 
Our interpretation of these results is presented in Section 5. 
Finally, in Section 6 we summarize our results, and suggest future tests 
of this hypothesis.

\section{Sample}

In this section, we describe how we construct a kinematically unbiased sample of 
low-mass main-sequence stars based on SDSS DR8.

\subsection{Target Selection}

The spectroscopic samples in SDSS, as well as in its stellar-specific
sub-surveys, the Sloan Extension for Galactic Understanding and
Exploration (SEGUE-1; \citealt{Yanny2009}) and SEGUE-2 (Rockosi et al., in
preparation) are selected (targeted) based on a series of photometric
(magnitude and color) and/or proper-motion cuts. In order to construct a
kinematically unbiased sample, we avoid using stars whose target
criteria includes any proper-motion cuts. Since the target criteria for
stars in the different surveys are not identical, we first examined
their detailed descriptions to determine which might be best used for
our present purpose. After some consideration, from SDSS we include only
those objects targeted as {\it BHB candidates} (a subset of these are
used for a separate test below). From SEGUE-1, we select those objects
whose target names are either {\it K-dwarf candidates}, {\it G-dwarf
candidates}, or {\it low-metallicity candidates}. We do not select any
objects from the SEGUE-2 sub-survey. The combined sample comprises the
basis for our subsequent analysis. Note that our sample is based
initially on the target criterion, not on how the star was classified
after its spectrum was obtained.

\subsection{Selection of G/K-type dwarfs with Reliable Stellar Parameters and 
Kinematics}

In order to construct a sample of low-mass, main-sequence dwarfs with
reliable estimates of atmospheric metallicity, [Fe/H], 3-D positions,
and space motions, we further restrict our selection to meet additional
criteria. In this process, we make use of the stellar parameters derived 
from the most recent version of the SEGUE Stellar Parameter Pipeline 
(SSPP; see \citealt{Lee2008a, Lee2008b, AP2008, Smolinski2011}). 
Note that the SSPP metallicity estimate is optimized for low-mass dwarfs, 
in essentially the same way as in \cite{Schlesinger2012}.

First, from the above-mentioned stellar sample, we select stars with
S/N$>20$ per 1 \;{\AA} pixel, colors in the range $0.48 < {(g-r)}_0 <
0.75$, derived surface gravity $\log g > 4.1$, and effective
temperatures in the range $4500\;{\rm K} < T_{\rm eff} < 6000\;{\rm
K}$. Here, ${(g-r)}_0$ is the reddening-corrected $(g-r)$ color based
on application of the prescription by \cite{Schlegel1998}. The lower
limit on the S/N ratio ensures that errors in [Fe/H] derived by the SSPP
are not too large 
(typically smaller than $0.15\;{\rm dex}$; \citealt{AP2008}). 
The lower limit on $\log g$ is set to reliably include dwarf stars. The
upper and lower limits on effective temperature $T_{\rm eff}$ correspond
to G0 and K4, respectively \citep{Habets1981}\footnote{Note that the MK
system is defined for solar-metallicity stars, so these ranges on
spectral type only loosely apply to low-metallicity stars.}. 
We then select those stars with one-sigma errors in line-of-sight
velocity, $v_{\rm los}$, smaller than $20 \; {\rm km \;s^{-1}}$, and
with one-sigma proper-motion errors motion smaller than 
$5\;{\rm mas\;{yr}^{-1}}$.

\subsection{Construction of a Volume-Limited Sample} \label{volume-limited sample}

In order to approximate a sample that reflects the nature of
a volume-limited sample, we retain stars with absorption-corrected $r_0$
magnitudes in the range $15 < r_0 < 18.45$. Then, we select those stars
with heliocentric distances, $d$, in the range 0.84 kpc $< d <$ 1.64 kpc,
and with atmospheric metallicities in the range $-2.0 <$[Fe/H]$< -0.5$, as
derived by the SSPP \citep{Beers2012}. These distance and metallicity
ranges are designed to ensure that the G- and K-type dwarfs fairly explore
the same volume (that is, we seek to avoid populating the volume differently
for the two spectral types), based on the set of 10 Gyr model isochrones
described in \cite{An2009}. Then, we divide our sample stars into G-
and K-dwarf samples based on $T_{\rm eff}$. Throughout this paper, those
dwarfs with $5250\;{\rm K} < T_{\rm eff} < 6000\;{\rm K}$ are referred
to as G-type dwarfs, which (for solar-abundance stars) corresponds to G0-G9
\citep{Habets1981}, while those with $4500\;{\rm K} < T_{\rm eff} \le
5250\;{\rm K}$ are referred to as K-type dwarfs, which correspond to K0-K4
(for solar-abundance stars). Our final sample consists of 7124 G-type
dwarfs and 3257 K-type dwarfs. 

The mean distances of the two samples, 
$1.263 \pm 0.003 \;{\rm kpc}$ (G-type dwarfs) and 
$1.254 \pm 0.004 \;{\rm kpc}$ (K-type dwarfs) are 
quite close to one another. A two-sample
Kolmogorov-Smirnov test of the distributions of heliocentric distance
for the G- and K-dwarf samples is unable to reject the null hypothesis
that they are drawn from the same parent distribution. The $p$-value of
the test ($0.081$) is larger than the widely adopted threshold of
$0.05$; we conclude that the spatial distributions for these two samples
do not significantly differ.

\subsection{Caveats on a Metallicity Selection Bias}

We are aware that our sample is affected by a selection bias in which
more metal-poor stars are preferentially observed (see section 4.7 of
\citealt{Schlesinger2012}). Such a selection bias must be treated with
care if we are concerned with extracting the metallicity distribution
functions for stars in our sample. However, this metallicity-dependent
selection bias should not impact our derivation of the median rotational
velocity ($V_\phi$ in a cylindrical system) for stars at a given {\it
fixed} metallicity, since it does not depend on {\it how many} sample
stars are used. In the following, we compare the relationship between
$\langle V_\phi \rangle_{\rm med}$ -- by which we indicate the median rotational velocity --
and [Fe/H] for different stellar subsamples.

\begin{figure}
\begin{center}
	\includegraphics[angle=-90,width=0.95\columnwidth]{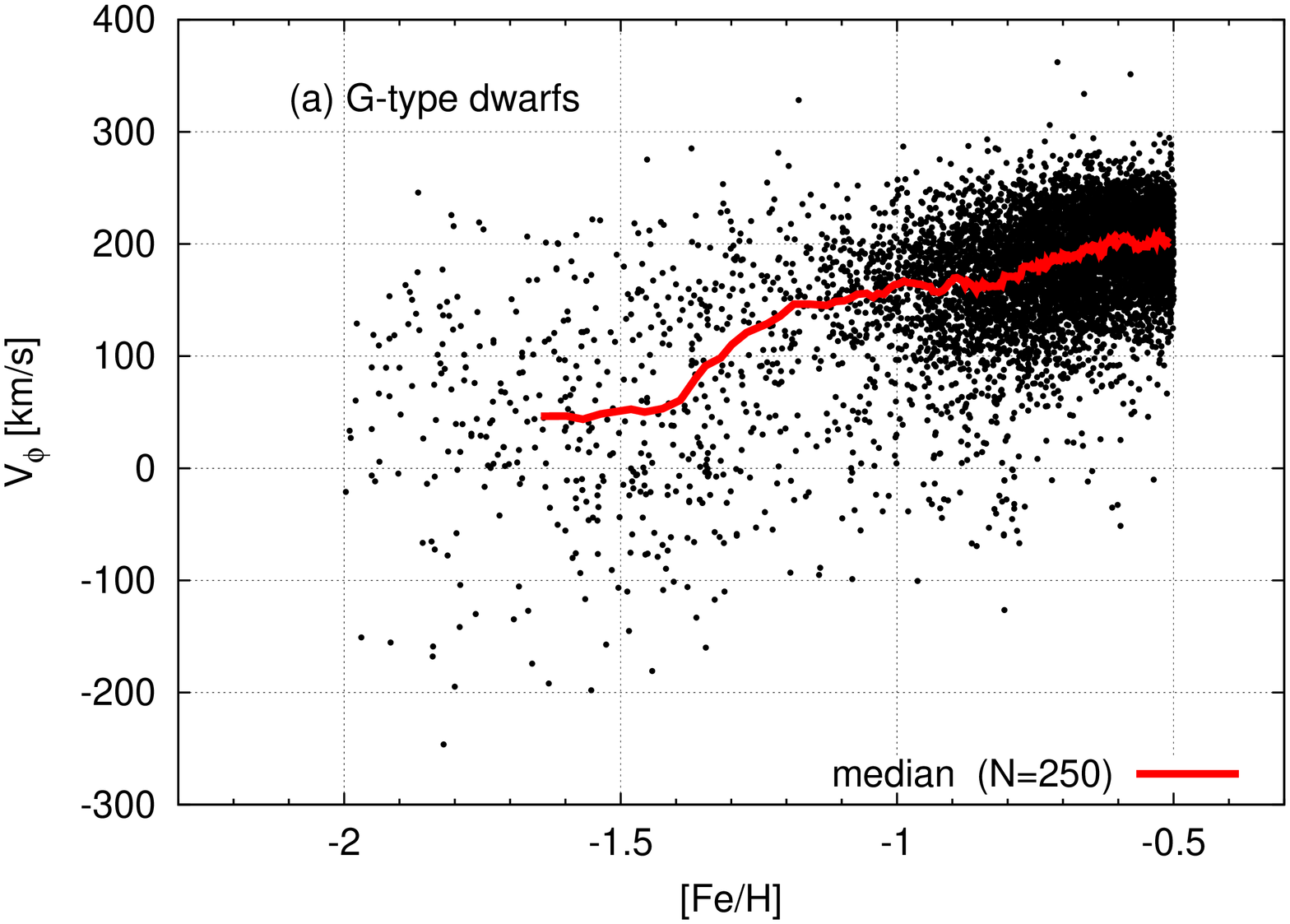} \\
	\includegraphics[angle=-90,width=0.95\columnwidth]{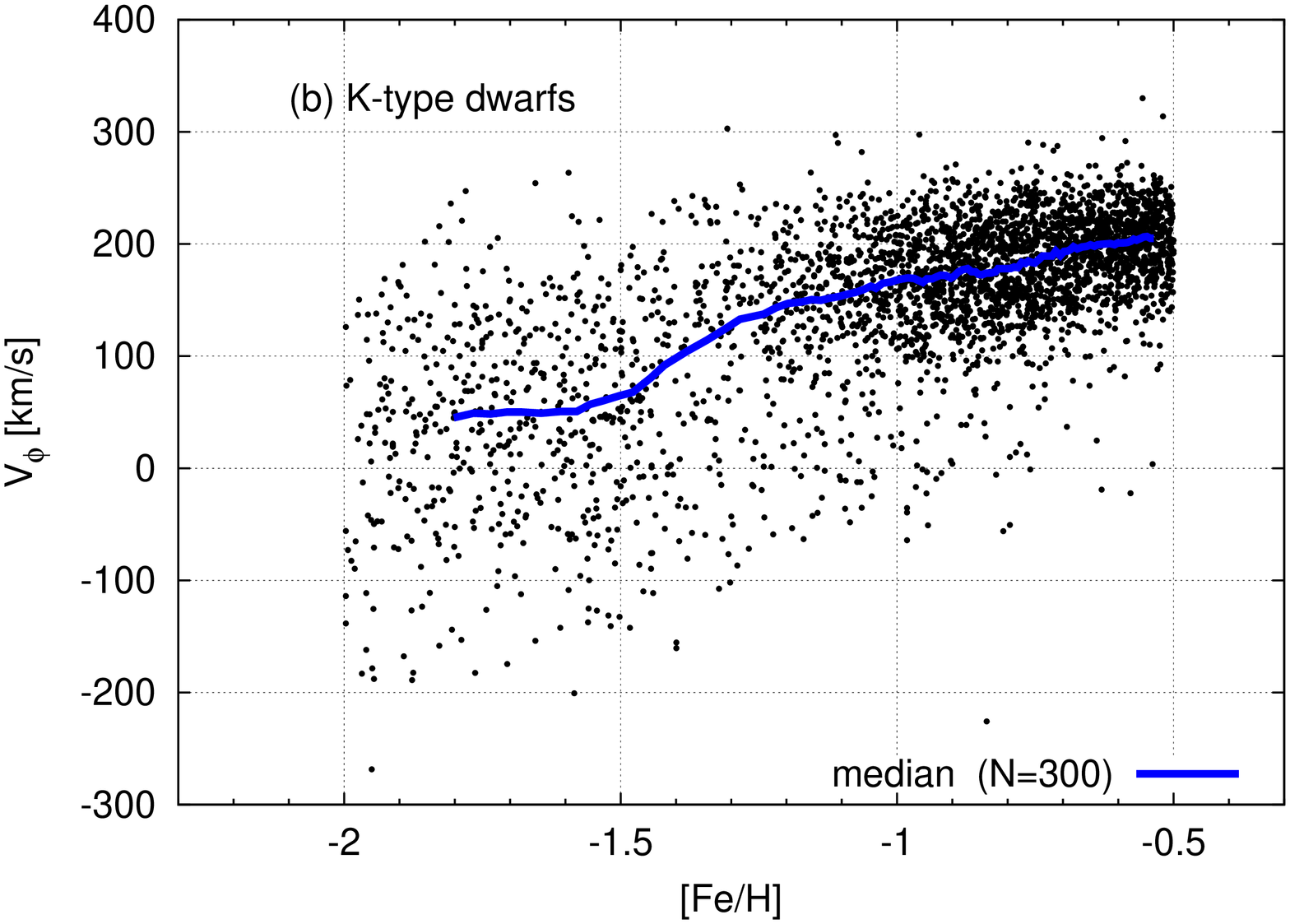} \\
	\includegraphics[angle=-90,width=0.95\columnwidth]{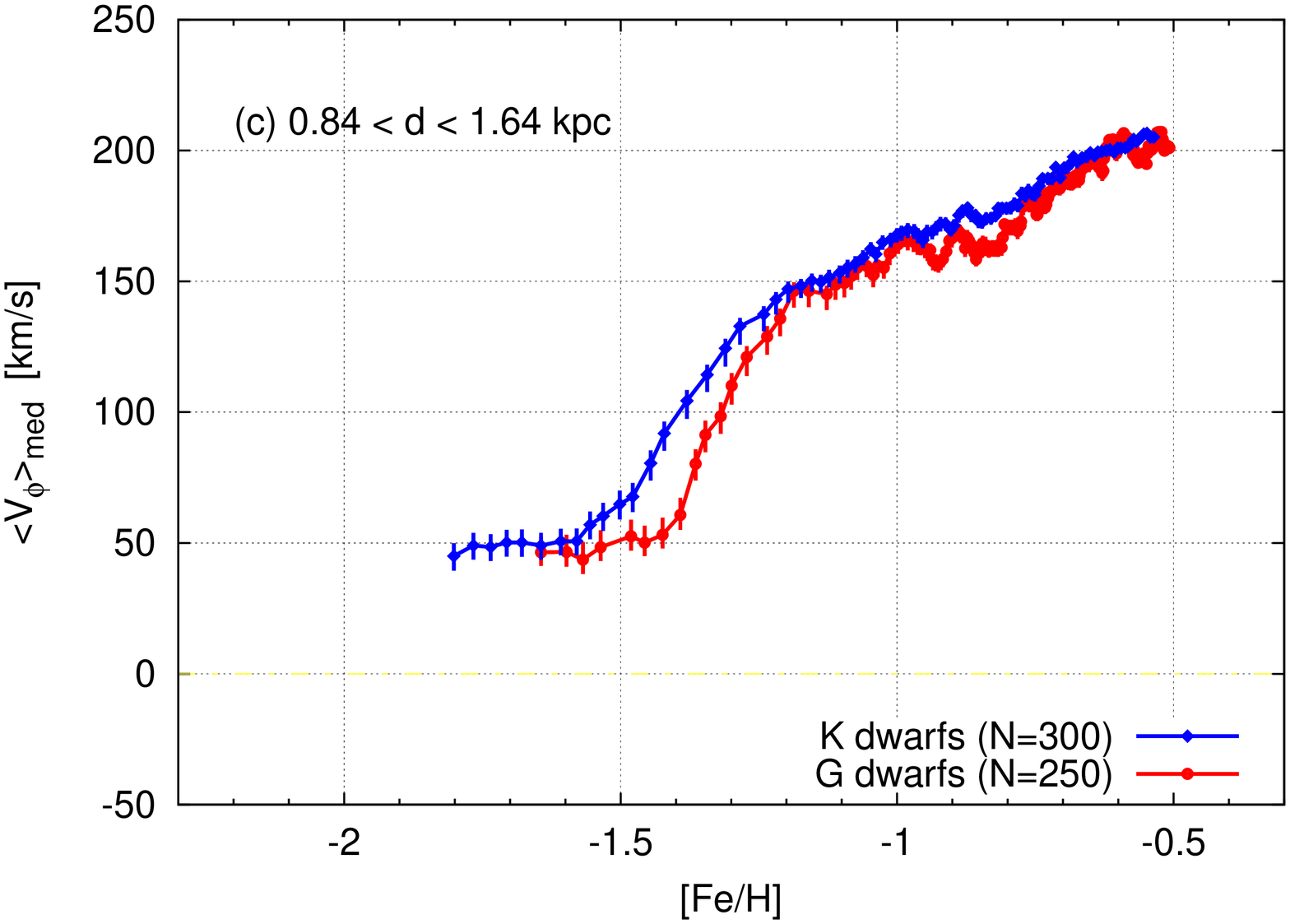} \\	
\end{center}
\caption{
Distribution of our sample stars in the $V_\phi$-[Fe/H] space and the
median value of $V_\phi$ for binned samples in the metallicity range $-2.0 < $ [Fe/H] $< -0.5$. (a) Results for G-type
dwarfs. We bin $N=$ 250 stars sorted in [Fe/H], moving through the
sample in steps of $N/10$ stars, and show the median value of $V_\phi$ (or $\langle V_\phi \rangle_{\rm med}$) 
at the median value of [Fe/H]. (b) Results for K-type dwarfs. The
analysis is the same as in (a), but adopting $N=300$ instead. (c)
Comparison of the median values $\langle V_\phi \rangle_{\rm med}$ for G- and K-type dwarfs. Error
bars in $\langle V_\phi \rangle_{\rm med}$ are estimated by assuming a
Gaussian-like distribution of $V_\phi$. }
\label{fig 1 pm}
\end{figure}

\begin{figure}
\begin{center}
	\includegraphics[angle=-90,width=0.95\columnwidth]{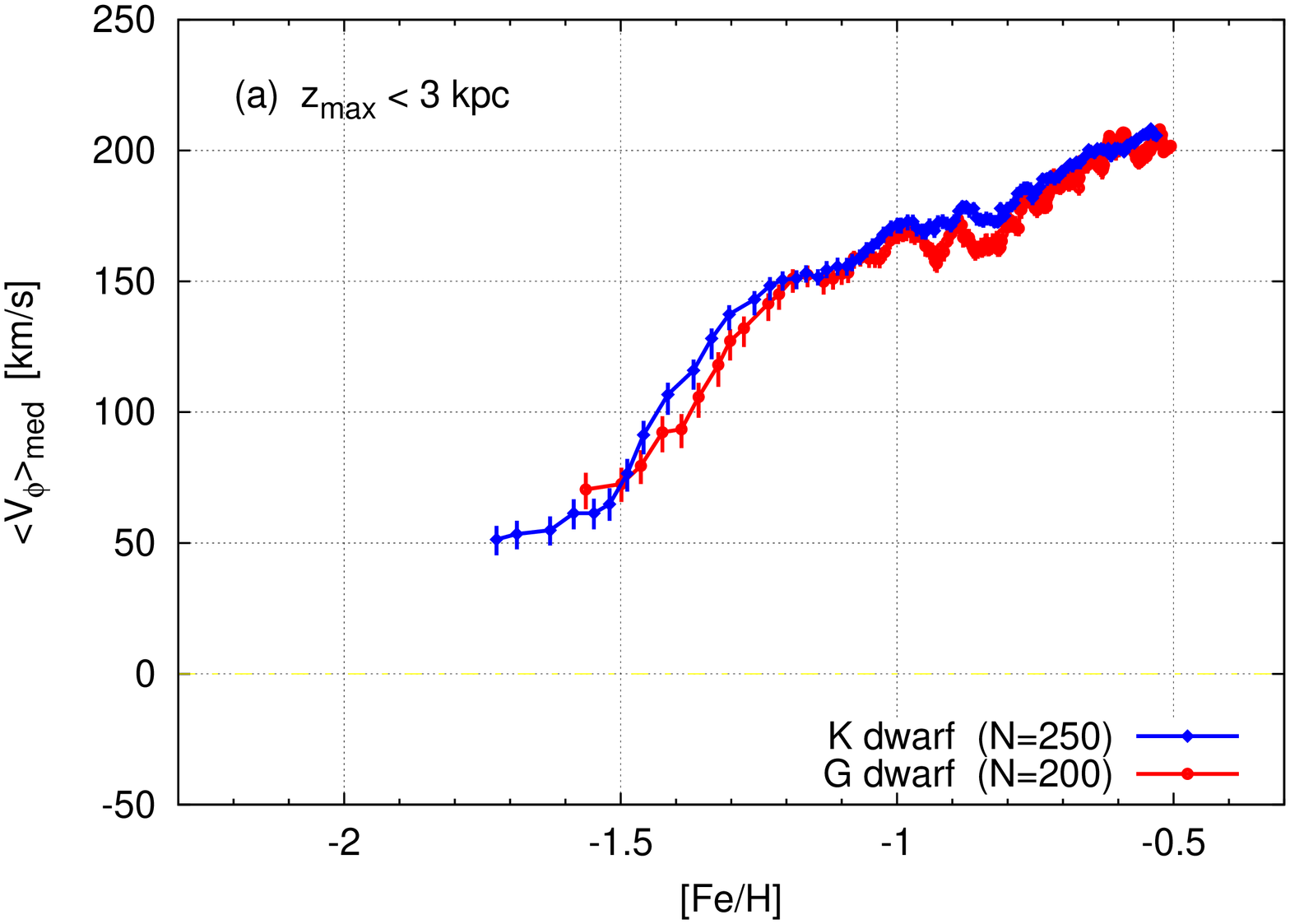} \\
	\includegraphics[angle=-90,width=0.95\columnwidth]{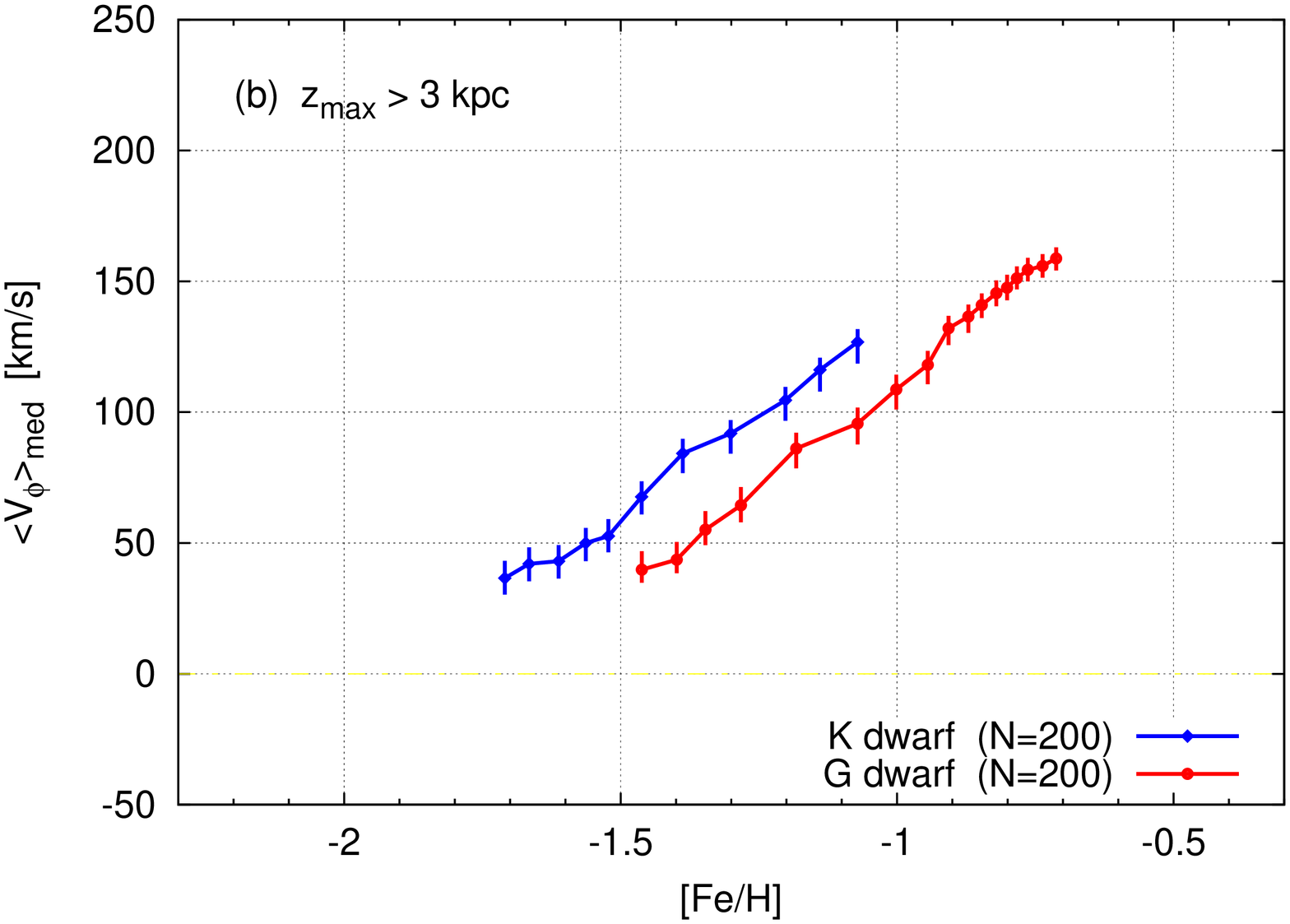} \\
\end{center}
\caption{
Median value of $V_\phi$ as a function of [Fe/H] for G-type dwarf (red) and
K-type dwarf (blue) subsamples having different ranges of $z_{\rm max}$
(maximum orbital excursion perpendicular to the Galactic disk plane).
The results for sample stars with $z_{\rm max}<$ 3 kpc and $z_{\rm max}>$
3 kpc are shown in panels (a) and (b), respectively. The binning 
procedure is the same as in Figure \ref{fig 1 pm}, and the adopted value
of bin size $N$ for each subsample is shown on the panel. }
\label{fig 2 zmax}
\end{figure}

\begin{figure}
\begin{center}
	\includegraphics[angle=-90,width=0.95\columnwidth]{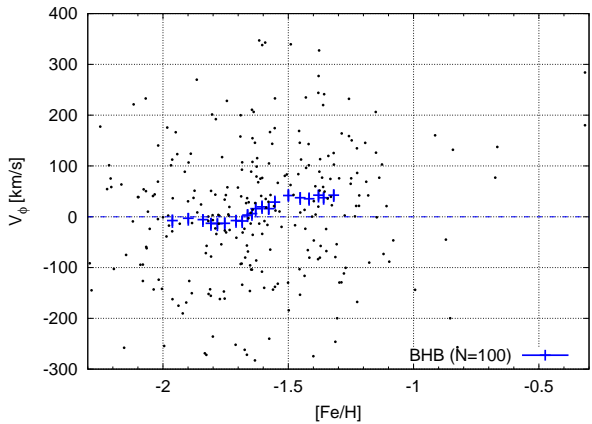}\\
\end{center}
\caption{
The distribution in the $V_\phi$-[Fe/H] space of BHB stars 
with $5\;{\rm kpc}<R<20\;{\rm kpc}$ and $2\;{\rm kpc}<|z|<5\;{\rm kpc}$. 
Also shown is the median value, $\langle V_\phi \rangle_{\rm med}$, for the binned sample. 
The binning procedure is the same as in Figure \ref{fig 1 pm}, 
but by binning $N=$ 100 stars and moving through each sample in steps of $N/10$ stars. 
The associated error bar on $\langle V_\phi \rangle_{\rm med}$ represents the uncertainty 
estimated by assuming a Gaussian-like distribution of $V_\phi$. 
}
\label{fig 3 BHB}
\end{figure}

\section{Median Rotational Velocity as a Function of Metallicity}\label{result}

\subsection{Kinematical Information}\label{Kinematical Information}

In this paper we assume that the Local Standard of Rest (LSR) is on a
circular orbit with a rotation speed of $220\;{\rm km\;s^{-1}}$
\citep{Kerr1986,Bovy2012}, and that the Galactocentric distance of the
Sun is $R_{\odot} = 8.5\;{\rm kpc}$ \citep{Ghez2008,Koposov2010}. We
also assume that the peculiar motion of the Sun with respect to the LSR
is $(U_\odot, V_\odot, W_\odot) = (10.0, 5.3, 7.2) \;{\rm km\;s^{-1}}$
\citep{Dehnen1998}. 

With these assumptions, and the available estimated distances,
line-of-sight velocities, and proper motions, we calculate the
3-D positions and space motions of our sample stars. Noting
that the fractional error in the SSPP distance is around 10-20\%, and
that the typical error in line-of-sight velocity and proper motion is
around 2 ${\rm km\;s^{-1}}$ and 3.5 ${\rm mas\;{yr}^{-1}}$, 
respectively, the associated error in each of the velocity
components of our sample stars is typically around 
30-40 ${\rm km\; s^{-1}}$. 

\subsection{Median Rotational Velocity Behavior with [Fe/H]}

Panels (a) and (b) of Figure \ref{fig 1 pm} show the distribution of G
and K-type dwarfs, respectively, in the $V_\phi$-[Fe/H] space. In each panel,
a solid line indicates the running median $\langle V_\phi \rangle_{\rm med}$ as a function of
[Fe/H], sweeping through the sample with overlapping bins of $N$ stars
sorted in [Fe/H] (with an overlap of $0.9 N$ stars per bin), using $N=$
250 G-type dwarfs and $N=300$ K-type dwarfs. Panel (c) of Figure 1 compares the
two distributions, with an associated error in $\langle V_\phi \rangle_{\rm med}$ estimated by
assuming a Gaussian-like distribution of $V_\phi$. For both the G- and
K-dwarf samples, we note a gradual transition from a nearly
non-rotating, halo-dominated region at [Fe/H]$<$ [Fe/H$]_{\rm knee}
\simeq -1.5$ to a rapidly-rotating, thick-disk dominated region at
[Fe/H]$\gtrsim -1.0$.\footnote{Note that most of our sample stars are
located more than 0.5 kpc away from the disk plane, while the scale
heights of thin- and thick-disk components are around 0.25-0.35 kpc and
0.7-1.2 kpc, respectively \citep{Yoshii1982, Gilmore1983, Yoshii2013}.}
However, we also note that there appears to be a systematic offset in
the $\langle V_\phi \rangle_{\rm med}$-[Fe/H] relation between the G- and K-type dwarfs, such
that the offset in [Fe/H] increases as $\langle V_\phi \rangle_{\rm med}$ decreases. The offset
in [Fe/H] (when $\langle V_\phi \rangle_{\rm med}$ is fixed) increases as a function of
$\langle V_\phi \rangle_{\rm med}$, from $\delta = 0.05$ dex at $\langle V_\phi \rangle_{\rm med} \simeq 150\;{\rm km\;
s^{-1}}$ to $\delta = 0.20$ dex at $\langle V_\phi \rangle_{\rm med} \simeq 50\;{\rm km\;
s^{-1}}$, below which the $\langle V_\phi \rangle_{\rm med}$-[Fe/H] relation becomes flat. In
other words, the offset becomes larger in the more metal-poor, and
therefore more halo-dominated region. 

It is also intriguing to note that the offset becomes milder at $\langle
V_\phi \rangle_{\rm med} > 150\;{\rm km\;s^{-1}}$, where the
contribution from the thick disk becomes larger, and that the offset
becomes invisible at $\langle V_\phi \rangle_{\rm med} > 170\;{\rm km\;
s^{-1}}$ or [Fe/H]$>-0.8$, where the sample is dominated by thick-disk
stars. 

We note here the possibility that some perturbative mechanisms (such as
close encounters with giant molecular clouds in the disk plane) could
result in a mass segregation of disk stars, and produce a difference in
the $\langle V_\phi \rangle_{\rm med}$-[Fe/H] relation. We check this
possibility using a simple mass distribution model for the Milky Way,
and find that the difference in the stellar masses of G- and K-type
dwarfs is too small to cause the observed offset of 30-40 ${\rm km\;
s^{-1}}$ at [Fe/H]$\simeq -1.4$. Therefore, hereafter we interpret the
detected offset in $\langle V_\phi \rangle_{\rm med}$-[Fe/H] relation as
an offset in [Fe/H], not as one in $\langle V_\phi \rangle_{\rm med}$.

Next, 
we derive the orbital parameters for each of the sample stars, 
assuming the St{\"a}ckel-type gravitational potential of the 
Milky Way adopted by \cite{CB2001}. 
We divide our G- and K-dwarf 
samples according to whether $z_{\rm max}$ is larger or smaller than 3 kpc 
($z_{\rm max}$ denotes the largest orbital excursion perpendicular to the
Galactic disk plane achieved by a given star during its orbit), 
and perform the same analysis as carried out above.

Panels (a) and (b) in Figure \ref{fig 2 zmax} show the median value of 
$V_\phi$, as a function of [Fe/H], for our sample stars with $z_{\rm
max}<$ 3 kpc and $z_{\rm max}>$ 3 kpc, respectively. The binning
procedure is the same as used for Figure \ref{fig 1 pm}, with the $N$
employed shown on each panel. Although the offset in $\langle V_\phi \rangle_{\rm med}$-[Fe/H] relation 
is not clear in panel (a), it becomes very apparent in panel (b). 
The magnitude of the offset on panel (b) is $\delta \simeq 0.20$ dex, 
which is as large as the maximum offset seen in Figure \ref{fig 1 pm}.

\section{Plausibility of Our Results}

In section \ref{result}, we find that G- and K-type dwarfs show
different behaviour in the $\langle V_\phi \rangle_{\rm med}$-[Fe/H]
relation. Here we discuss how the selection bias in SDSS or
observational errors may affect our results and investigate the
plausibility of our results.

\subsection{Selection Bias in SDSS}

\cite{Schlesinger2012} point out that, 
even if they take into account the metallicity-dependent selection bias
inherent in SDSS/SEGUE, the fraction of metal-poor K-type dwarfs is
larger than that of G-type dwarfs, especially in the high $|z|$-region
($|z|$ is the distance from the Galactic disk plane). In order to
investigate if this discrepancy is involved with producing the observed
offset in the $\langle V_\phi \rangle_{\rm med}$-[Fe/H] relation, we
select those sample stars that appear also in the \cite{Schlesinger2012}
sample, and check the $|z|$-dependence of the $\langle V_\phi
\rangle_{\rm med}$-[Fe/H] relation. We find that the offset can be
confirmed independent of $|z|$, which suggests that the observed offset
is not relevant to the discrepancy found in \cite{Schlesinger2012}.

\subsection{Observational Errors}

\subsubsection{Systematic Errors in SSPP Metallicity Estimates}

The simplest explanation of our finding that G- and K-type dwarfs show
different $\langle V_\phi \rangle_{\rm med}$-[Fe/H] relations is that it
is due to a temperature-related systematic error in the determination of
[Fe/H] by the SSPP. If such a systematic error in [Fe/H] exists, we may
detect a systematic difference between SSPP metallicity and the
metallicity derived from high-resolution spectroscopy for G/K-type
dwarfs. We check this possibility by using five G-type dwarfs and two
K-type dwarfs taken from Table 4 of \cite{AP2008}, and find that G-type
dwarfs tend to have $\sim$ 0.2 dex higher SSPP metallicity than the
high-resolution metallicity, while the two estimates of metallicity more
or less agree with each other for K-type dwarfs.\footnote{
According to
\cite{AP2008}, possible systematic errors in [Fe/H] for SDSS/SEGUE
spectra with S/N $>20$ are less than 0.15 dex. However, this value does
not directly apply to our sample dwarfs, since their sample includes not
only dwarfs but also more luminous giants. Indeed, their Table 4
indicates that the uncertainty in the SSPP metallicity for G/K-type
dwarfs might be as large as 0.3 dex. } 
Although the sample size (in total seven) 
is insufficient to be confident, the uncertainty in SSPP metallicity
might influence the observed offset in $\langle V_\phi \rangle_{\rm
med}$-[Fe/H] relations in Figure \ref{fig 1 pm} or \ref{fig 2 zmax}.  

However, if the apparent offset in Figure \ref{fig 1 pm} is due to a
spectral-type dependent systematic error in the SSPP metallicity, we
would also expect a similar offset in Figure \ref{fig 2 zmax}(a). The fact that
we see a clear offset only in Figure \ref{fig 1 pm} and \ref{fig 2
zmax}(b) and not in Figure \ref{fig 2 zmax}(a), may indicate that the
existence or non-existence of the offset in the $\langle V_\phi
\rangle_{\rm med}$-[Fe/H] relation in these figures is real. This notion
is supported by the mock data analysis presented in section \ref{mock}.

\subsubsection{Random Errors in Proper Motion}

The largest contribution to the uncertainty in the measured velocity
components of our sample stars arises from the observational errors in
proper motion, which are typically as large as 3.5 ${\rm mas\;
{yr}^{-1}}$. However, the proper-motion error does not seem to affect
the median value, $\langle V_\phi \rangle_{\rm med}$, since it is a
purely random error in most cases. In order to estimate the effect of
proper-motion errors on our results, we first construct 100 error-added
samples of G- and K-type dwarfs. To do this, we randomly add/subtract an
error term to the observed proper motion that obeys a Gaussian
distribution with standard deviation corresponding to the one-sigma
observational error in the proper motion. Then, we perform the same
analysis to these error-added samples, and compare the results with that
of the as-observed sample. After these calculations, we confirm that the
proper-motion error scarcely affects our results shown in Figure 
\ref{fig 1 pm} or \ref{fig 2 zmax}.

\subsubsection{Systematic Errors in SSPP Distance Estimates}

We next consider how systematic errors in the determination of distance
could affect our result. In our analysis, we select stars with distances
in the range 0.84 kpc $< d <$ 1.64 kpc. If the distances to K dwarfs are
over-estimated by the SSPP, then more K-type dwarfs in our sample are
expected to reside closer to the Galactic disk plane, and the fraction
of K-type dwarfs that belong to the thick disk is over-estimated in our
sample. In this case, we unintentionally compare the low-$|z|$ K dwarfs
and high-$|z|$ G-type dwarfs, leading to a higher value of $\langle
V_\phi \rangle_{\rm med}$ for K-type dwarfs at a fixed [Fe/H]. In order
to test this bias, we construct two different samples of K-type dwarfs
selected by the same criteria as in Section 2, but assigned distances
that are 20\% smaller or larger compared to the adopted distance. In
both instances, we confirm that the difference between the resultant
$\langle V_\phi \rangle_{\rm med}$-[Fe/H] relation and the original one
is less than $20\;{\rm km\;s^{-1}}$ at [Fe/H]$> -1.5$. We conclude that
a systematic error of $\sim$ 20\% in distance determination cannot
explain the observed difference in the $\langle V_\phi \rangle_{\rm
med}$-[Fe/H] relationship between the G- and K-type dwarfs.

\subsubsection{Mock Data Analysis} \label{mock}

In order to check the validity of our results, we also examine how
random and/or systematic errors in the observed quantities (such as
distance and metallicity) could affect our results in Figures \ref{fig 1
pm} and \ref{fig 2 zmax}. For this purpose, we construct a set of
realistic mock catalogs in which the information on effective
temperature, surface gravity, metallicity, distance, Galactic latitude
and longitude is identical to that of our real sample,
\footnote{Since the values of $T_{\rm eff}$, $\log g$,
[Fe/H], and $d$ are the same as those of the real sample, the spatial and
metallicity distributions of G- and K-type dwarfs of any given mock
catalog are identical to those of our real sample. } but the
information on line-of-sight velocity and proper motion reflects a given
distribution function model as well as assumed realistic errors in
metallicity, distance, line-of-sight velocity and proper motion. Note
that the effect of metal accretion is not taken into account in our mock
catalogs. (See Appendix for the full description of our mock catalogs.)

We consider various models of random and/or systematic errors in the
observed quantities, and prepare 100 mock catalogs for each error model.
Then we perform the same analyses on our mock catalogs as applied to our
real sample, and derive the $\langle V_\phi \rangle_{\rm med}$-[Fe/H]
relations which correspond to Figures \ref{fig 1 pm} (full sample) and
\ref{fig 2 zmax} ($z_{\rm max}$-limited subsamples). 

First, we find that the $\langle V_\phi \rangle_{\rm med}$-[Fe/H]
relations for G- and K-type dwarfs in our mock catalogs are
statistically identical, if we do not introduce any spectral-type
dependence in the observational errors. Noting that the spatial
distributions of the mock G- and K-type dwarfs are identical to those of
the real G- and K-type dwarfs, this result justifies our starting point
that the spatial distributions of our G- and K-type dwarfs are
essentially identical (section \ref{volume-limited sample}). 

Secondly, we find that the offset in the $\langle V_\phi \rangle_{\rm
med}$-[Fe/H] relation similar to that in Figure \ref{fig 1 pm} is seen
only when the random or systematic errors in [Fe/H] depend on the spectral
type. In addition, we find that when such an offset is seen in the full
sample, a similar offset is also seen in the low-$z_{\rm max}$ subsample,
and vice versa. In other words, random or systematic errors in [Fe/H]
either produce an offset for {\it both} Figures \ref{fig 1 pm}(c) and \ref{fig
2 zmax}(a), or produce no offset for {\it both} of these figures. This finding
is in sharp contrast with our results for the real sample, in which a
clear offset is seen only in Figure \ref{fig 1 pm}(c) and not in Figure
\ref{fig 2 zmax}(a). Moreover, we also find that the offset in the
$\langle V_\phi \rangle_{\rm med}$-[Fe/H] relation for the high -$z_{\rm
max}$ subsample is, if anything, less clear than that for the low-$z_{\rm
max}$ subsample. Presumably, this result is due to the smaller number of stars in
the high-$z_{\rm max}$ subsample. This finding is also opposite to our
results for the real sample, in which a clearer offset is seen in Figure
\ref{fig 2 zmax}(b) than in Figures \ref{fig 2 zmax}(a). 

These results suggest that the observed offset in the $\langle V_\phi
\rangle_{\rm med}$-[Fe/H] relation cannot be explained by observational
errors that may or may not depend on the spectral type. Therefore, we
conclude that the observed offset in Figures \ref{fig 1 pm} and \ref{fig
2 zmax}(b), as well as apparent non-existence of the offset in Figure
\ref{fig 2 zmax}(a), are all real.  

\section{Discussion}

In this paper we have derived the relation of $\langle V_\phi
\rangle_{\rm med}$ (median value of $V_\phi$) as a function of [Fe/H]
for halo G- and K-type dwarfs from SDSS DR8. We find that the run of
$\langle V_\phi \rangle_{\rm med}$ vs. [Fe/H] can be characterized by a
boundary metallicity, [Fe/H$]_{\rm knee}$, below which the nearly
non-rotating halo stars dominate, and that [Fe/H$]_{\rm knee}$ for
G-type dwarfs occurs at a higher abundance than that for K-type dwarfs
by an offset of $\delta \simeq 0.20\;{\rm dex}$. This offset is also
seen for those sample stars that have large vertical motion. Below we
consider the implications of this result.

\subsection{Evidence of Metal Accretion onto Main-Sequence Halo Stars} \label{evidence}

We interpret this non-zero value of $\delta \simeq 0.20\;{\rm dex}$
between the characteristic metallicities of halo G- and K-type dwarfs to
be the first tentative evidence that halo stars have been {\it
externally} polluted by the accretion of metal-enriched gas from their
natal clouds. The effect is more noticeable for G-type dwarfs, whose
convective envelopes are shallower than those of K-type dwarfs, since
they will not have diluted accreted material as fully.

If this interpretation is correct, we expect an even larger offset in
metallicity for red giants (e.g., K giants), which possess deeper
surface convective envelopes than K-type dwarfs \citep{Yoshii1981}. That
is, the knee of the $\langle V_\phi \rangle_{\rm med}$-[Fe/H] relation
for K giants, if plotted as in Figure \ref{fig 1 pm}, should be located
to the left of that for K dwarfs. Although there is no large publicly
available database of nearby K giants, other types of post-main-sequence
stars might serve as alternatives, since the surface heavy element
abundance is expected to remain unchanged when red giants evolve into
either horizontal-branch stars or RR Lyrae stars. During the course of
this evolution, red giants lose mass from their surface convective
envelopes. The total amount of mass loss is expected to be $\simeq0.2\;
M_\odot$ for a red giant of initially $0.8\;M_\odot$ (see
\citealt{Yoshii1981}, and references therein), which is smaller than the
total mass of the surface convective envelope ($\simeq0.3\;M_\odot$) in the
red-giant stage \citep{Sweigart1978}. 

Thus, instead of red giants, we have explored the $\langle V_\phi
\rangle_{\rm med}$-[Fe/H] relation for 290 blue horizontal-branch (BHB)
stars with $T_{\rm eff} > 7500\;{\rm K}$, $\log g <3.8$, located at $2\;
{\rm kpc}<|z|<5\;{\rm kpc}$, $5\;{\rm kpc}<R<20\;{\rm kpc}$, taken from
\cite{Xue2011}. Here, $R$ and $|z|$ are the Galactocentric distance
projected onto the Galactic disk plane and the distance above or below
the disk plane, respectively. The cuts in $T_{\rm eff}$ and $\log \; g$
are adopted following the selection criteria proposed by R. Santucci et
al. (in preparation). 

In Figure \ref{fig 3 BHB}, we show the distribution of BHB stars in the
$V_\phi$-[Fe/H] space, along with the median value $\langle V_\phi
\rangle_{\rm med}$ for the binned sample. From inspection of this
figure, the $\langle V_\phi \rangle_{\rm med}$ of BHB stars is nearly
constant (consistent with 0 ${\rm km\;s^{-1}}$) at [Fe/H] $<$
[Fe/H$]_{\rm knee} \simeq -1.7$.\footnote{
The plateau value of $\langle V_\phi \rangle_{\rm med}$ at [Fe/H] $<$
[Fe/H$]_{\rm knee}$ for the BHB sample ($\simeq 0\;{\rm km\;s^{-1}}$) is
lower than that for G/K-type dwarfs ($\simeq 50\;{\rm km\;s^{-1}}$; see
Figure \ref{fig 1 pm}). This discrepancy seems to arise from the different
spatial regions covered by these samples. In fact, \cite{CB2000} find
that the mean rotational velocity $\langle V_\phi \rangle$ for nearby metal-poor stars
with [Fe/H] $< -1.5$ is higher for those sample stars with smaller $|z|$
[see their Figure 3(a)], which is consistent with our discrepancy. }
This suggests that inner-halo BHB stars dominate over metal-weak
thick-disk BHB stars below [Fe/H] $=$ [Fe/H$]_{\rm knee}$ at $2\; {\rm
kpc}<|z|<5\; {\rm kpc}$. Noting that a lower value of [Fe/H$]_{\rm
knee}$ is expected for a sample of stars with lower $|z|$ (due to the
larger fraction of thick-disk stars in the sample), we expect that
[Fe/H$]_{\rm knee}$ for BHB stars would be even lower than $-1.7$ if we
could obtain a BHB sample with distances in the range 0.84 kpc $< d <$
1.64 kpc, as was used above for the G- and K-dwarf samples. Since the
positions of [Fe/H$]_{\rm knee}$ for the G and K-type dwarfs are $-1.4$
and $-1.6$, respectively, in this distance range (see Figure \ref{fig 1
pm}), it follows that [Fe/H$]_{\rm knee}$ for BHB stars should be lower,
by {\it at least} 0.3 dex and 0.1 dex, respectively, if these samples
could be fairly compared. This may indicate that the BHB stars are even
less affected by surface metal pollution than K-type dwarfs. The sample
of 290 BHB stars is not sufficiently large to be certain of this effect,
but the present result does serve to support the accretion hypothesis
for halo stars. Surveys such as LAMOST and Gaia will provide larger
kinematically unbiased samples of low-mass dwarfs and red giants. At that
stage, we will be able to compare the chemo-dynamical correlations of
these stars more rigorously, and test the metal accretion hypothesis
thoroughly.

\subsection{Where did the Metal Accretion Take Place?}

As mentioned in the Introduction, the efficiency of metal accretion in
halo stars is proportional to $v_{\rm rel}^{-3}$, where $v_{\rm rel}$ is
the relative velocity of the star and the colliding gas. Therefore,
metal accretion onto halo stars may or may not be important, depending
on the nature of the environments in which it could take place. Recent
observations \citep{Carollo2007, Carollo2010, Beers2012} and numerical
simulations \citep{Font2011, McCarthy2012, Tissera2013} suggest that the
stellar halo of the Milky Way comprises two distinct components with
different origins, which are often referred to as the inner- and
outer-halo populations. In this picture, inner-halo stars formed from
relatively more massive sub-galactic systems ($10^{9-10} M_\odot$) in
the main progenitors of the Milky Way, while much smaller sub-galactic
systems ($10^{7-8} M_\odot$ or less), similar to lower-mass dwarf-like galaxies,
were disrupted to contribute the bulk of the low-metallicity outer-halo
stars. In the progenitor systems of the inner-halo stars, the internal
velocity dispersion is expected to be $\sim 30\;{\rm km\; s^{-1}}$
\citep{Tissera2013}, so that the above-mentioned $v_{\rm rel}$ may have
been too large for inner-halo stars to experience efficient metal accretion. 
On the other hand, the internal velocity dispersion of 
ultra-faint dwarf galaxies 
-- whose constituent stars are typically as metal-poor as [Fe/H] $\lesssim -2.0$, 
similar to nearby outer-halo stars -- 
is $\sim 5\;{\rm km\; s^{-1}}$ (\citealt{Simon2007, Walker2009}; see also \citealt{YoshiiArimoto1987}), 
so that $v_{\rm rel}$ may have been small enough to enable efficient metal accretion onto constituent stars. 
Therefore, in this inner/outer halo picture, the outer-halo stars
are much more likely to have experienced metal accretion. In other
words, if the dual halo picture is correct, we expect that inner-halo
dominated sample of G- and K-type dwarfs would exhibit little or no
offset in the $\langle V_\phi \rangle_{\rm med}$-[Fe/H] relation, while
the outer-halo dominated sample of G- and K-type dwarfs would show
noticeable offsets. 

Observationally, it is suggested that the fraction of outer-halo stars
increases as $z_{\rm max}$ increases \citep{Carollo2010}. 
Thus, if
we divide each of our G- and K-dwarf samples into two subsamples
with respect to $z_{\rm max}$, the fraction of outer-halo stars is
expected to be larger in the higher-$z_{\rm max}$ subsample than in the
lower-$z_{\rm max}$ one. It follows that we expect a clearer offset in
the $\langle V_\phi \rangle_{\rm med}$-[Fe/H] relation for the
higher-$z_{\rm max}$ subsample. 
This expectation is indeed consistent with the results in Figure
\ref{fig 2 zmax}, in which we observe a clearer offset in the $\langle
V_\phi \rangle_{\rm med}$-[Fe/H] relation for our sample stars with
$z_{\rm max} > 3 \; {\rm kpc}$, and not for those stars with $z_{\rm max}
< 3 \; {\rm kpc}$.\footnote{ Our boundary at $z_{\rm max} =$ 3 kpc is
smaller than that adopted in \cite{Carollo2010}, so that we have a
sufficient number of stars in the higher-$z_{\rm max}$ subsample.
However, this difference is not crucially important, since our aim is to
examine two subsamples with different fractions of inner/outer-halo
stars. Our higher-$z_{\rm max}$ subsample includes numerous stars with
$z_{\rm max}$ as large as 10-50 kpc. } These results can be understood
if a dual inner/outer halo applies to the Milky Way, as originally
suggested by \cite{Carollo2007}, and metal accretion is efficient only
for stars associated with the progenitors of the outer halo.

We note here that typical halo stars with $z_{\rm max}>$ 3 kpc would
penetrate through the Galactic disk with a velocity of $v_{\rm rel} >
100\;{\rm km\;s^{-1}}$ relative to the gas clouds in the disk.
Therefore, metal accretion when the halo stars penetrate the Galactic
disk plane accounts for only a negligible fraction of the total amount
of metals accreted onto halo stars, supporting the view of
\cite{Frebel2009}. However, the fact that we see a clear offset in panel
(b) of Figure 3 suggests that the outer-halo stars are likely to have
experienced metal accretion within their progenitor systems, which were
{\it later} disrupted.

\subsection{Impact on the Metallicity Distribution Function in the 
Stellar Halo System} 

One obvious impact of our results is that the metallicity distribution
function (MDF) of halo main-sequence stars may need to be re-examined.
In the hierarchical galaxy formation scenario, halo stars form in
sub-galactic systems that are later disrupted by tidal interaction with
the Galaxy, hence halo stars originating from these disrupted systems
are likely to have experienced metal accretion within these systems
\citep{Shigeyama2003, Suda2004, Komiya2010}. As a consequence, the
shape of the MDF of halo stars is expected to be skewed toward higher
[Fe/H] when compared to their original MDF, because the surface
metallicity is enhanced for stars with lower [Fe/H] due to their shallow
surface convective envelopes, while it is less enhanced for stars with higher
[Fe/H] due to their deeper surface convective envelopes \citep{Mengel1979}. 

In this respect, it is intriguing to note the apparent discrepancy
between the MDFs of main-sequence turn-off (MSTO) stars and BHB stars in
the outer halo region of the Milky Way. \cite{Sesar2011} show that the
median metallicity of MSTO stars at Galactocentric distances in the
range of $10\;{\rm kpc}<r<30\;{\rm kpc}$ is [Fe/H]$\simeq -1.5$, almost
independent of $r$.\footnote{ See panels W3 and W4 in Figure 12 of
\cite{Sesar2011}, in which the influence from known substructures in the
halo is not significant.} On the other hand, \cite{Beers2012} show that
the median metallicity of BHB stars at $10\;{\rm kpc}<r<40\;{\rm kpc}$
is [Fe/H]$\simeq-2.0$, almost independent of $r$.\footnote{ See the
right-hand panels in Figure 15 of \cite{Beers2012}, in which the
contamination from the Sagittarius dwarf galaxy is minimized, and Figure
4 of \cite{Carollo2007}. } This discrepancy can be well explained by the
accretion hypothesis. MSTO and BHB stars are of similar age, yielding no
systematic difference in the mass accreted in their main-sequence stage.
Thus, the surface metallicity of BHB stars is much less enhanced than
that of MSTO stars, because the red giants that were
progenitors of the stars presently on the BHB had very deep surface
convective envelopes compared to MSTO stars (see section
\ref{evidence}).

Furthermore, noting that the surface convective envelopes of MSTO stars
are shallower than those of G-type dwarfs, it is also interesting to note that
the apparent offset in the median metallicity of the MDF for MSTO and BHB
stars ([Fe/H]$_{\rm median, MSTO} - $ [Fe/H]$_{\rm median, BHB} \simeq 0.5 
\;{\rm dex}$) is consistent with the offset in the break metallicity
([Fe/H$]_{\rm knee, G} - $[Fe/H$]_{\rm knee, BHB} \gtrsim 0.3$ dex) of
the $\langle V_\phi \rangle_{\rm med}$-[Fe/H] relation for G-type dwarfs and BHB stars (see
section \ref{evidence}). Our interpretation of this consistency in terms
of the accretion hypothesis may be tested based on a much larger sample
of BHB stars or red giants that should be available in the near future.  

Another intriguing aspect of the observed halo MDF is the possible
existence of a cutoff metallicity at [Fe/H]$_{\rm cutoff}\; \simeq -4$,
below which the MDF shows a sharp decline (\citealt{Schorck2009,
Li2010}; see also \citealt{Yong2013}). There are not yet a sufficient
number of stars known with [Fe/H] $< -3.5$, let alone with [Fe/H] $<
-4.0$, to evaluate whether the cutoff is real, or simply the result of
small number statistics in the ultra metal-poor regime. Future
observations of G/K-type dwarfs might well detect a spectral-type dependence
of this cutoff metallicity, in the sense that [Fe/H]$_{\rm cutoff, K} <
$ [Fe/H]$_{\rm cutoff, G}$, if metal accretion and subsequent surface
metal enhancement does indeed take place among metal-poor halo stars.

\subsection{On the Case of $\omega$ Centauri}

Some globular clusters appear to be ideal places to test the metal 
accretion hypothesis, if they are regarded as closed systems and the 
constituent stars are coeval, or nearly so. 
\cite{Stanford2007} investigate the metallicities of stars in the 
globular cluster $\omega$ Cen, 
and find that the peak metallicity of near-turn-off stars is 
systematically higher than that of red giant branch (RGB) stars, 
by 0.01-0.05 dex (see their Figure 2). 
This offset is smaller than the uncertainty in 
the derived [Fe/H] for their sample (0.15-0.20 dex). 
However, 
taking into account that the surface convective envelope is 
deeper for RGB stars than near-turn-off stars, 
the sense of the reported gap is in agreement with the 
prediction from the metal accretion hypothesis.

If the reported gap (0.01-0.05 dex) between near-turn-off stars 
and RGB stars is due to metal accretion, we expect that the typical
metallicities of G- and K-type dwarfs in $\omega$ Cen are different from 
each other by less than 0.05 dex, 
because of the milder difference in 
the depth of surface convective envelope between G- and K-type dwarfs. 
If we take into account the observed offset of $\simeq 0.20$ dex in
Figure \ref{fig 2 zmax}(b), the metal accretion in typical progenitor
systems of the Milky Way outer halo is more significant than that in
$\omega$ Cen. Noting that the efficiency of metal accretion is
proportional to $v_{\rm rel}^{-3}$, the seemingly low efficiency within
$\omega$ Cen indicates that typical progenitor systems of the outer halo
must have smaller internal velocity dispersions than that of $\omega$
Cen. It is suggested that such progenitor systems are similar to the
currently observed ultra-faint dwarf galaxies 
(with central velocity dispersion $\sim 5 \; {\rm km\;s^{-1}}$), 
and that the contribution from large and
compact globular clusters (such as $\omega$ Cen, with central velocity
dispersion $\simeq 17 \; {\rm km\;s^{-1}}$;
\citealt{Sollima2009}) to the outer halo is not significant.

\subsection{Implication for the First Stars}

If primordial (zero-metal) stars from the very first generation (with
main-sequence masses below $0.8 M_\odot$) were able to form, they could
survive until today, and possibly be observed.  Here we consider this
possibility from the perspective of the accretion hypothesis. 

The mass range of the first stars has been long debated, but some
nucleosynthesis constraints have recently been placed by the discovery
of four hyper metal-poor (HMP; technically, Fe-poor) stars with
[Fe/H]$\lesssim-5$ (HE~01072-5240: \citealt{Christlieb2002};
HE~1327-2326: \citealt{Frebel2005}; HE~0557-4840: \citealt{Norris2007};
SDSS~J102915+172927: \citealt{Caffau2012}). Three of the four stars
(excluding SDSS~J102915+172927) exhibit elemental abundance patterns
that are significantly enhanced with carbon (C) and nitrogen (N), and
can be well-explained if they are second-generation stars, formed out of
an interstellar medium that was chemically enriched by supernova ejecta
from $25 M_\odot$ primordial stars \citep{Umeda2003, Iwamoto2005,
Nomoto2013}. 

However, if we allow for the metal accretion hypothesis, as suggested to apply
in this paper, it is possible to regard the above-mentioned HMP stars as
surviving primordial stars whose surfaces were polluted by the accretion
of the supernova ejecta of $25 M_\odot$ primordial stars
\citep{Shigeyama2003}. Since the mass of these stars is as small as
0.8$M_\odot$, our interpretation has an implicit requirement for the
mass of primordial stars to range from less than $1 M_\odot$ up to more
than a few tens of $M_\odot$, as predicted by some authors
\citep{Yoshii1986, Nakamura2001}. In this respect, it is worth
mentioning that SDSS~J102915+172927 is an 0.7$M_\odot$ star with no
significant C and N enhancement. The mass fraction of heavy elements
derived for this star, $Z<7.4\times10^{-7}$, is the lowest known to
date, and heavy-element cooling in metal-poor gas with $Z<Z_{\rm crit} =
10^{-5} - 10^{-6}$ is superseded by purely atomic or molecular hydrogen
or zero-metal cooling \citep{Silk1977, Yoshii1980, Bromm2001, Smith2009, Safranek2010}. 
Therefore, the discovery of a 0.7$M_\odot$ star with $Z <
Z_{\rm crit}$ indicates that this star could indeed have formed by
zero-metal cooling. In other words, the zero-metal cooling process turns
out to be able to produce primordial stars with the masses below $1
M_\odot$. 

We propose that at least some of the HMP stars could be surviving
primordial stars that have experienced surface metal pollution after
they formed. Given that the environments in which such pollution can
occur is limited, we expect that even more HMP metal-poor stars similar
to SDSS~J102915+172927, with $Z \ll Z_{\rm crit}$, may well be discovered.

\section{Summary} 

In this paper we have described possible observational evidence for the
surface metal pollution of halo stars due to the accretion of
metal-enriched material onto stellar surfaces, as theoretically
predicted by \cite{Yoshii1981}. If we take at face value our analysis of
an additional sample of BHB stars, it might be the case that the initial
metallicities of halo G-type dwarfs and halo K-type dwarfs are at least
$\sim$0.3 dex and $\sim$0.1 dex {\it lower} than their observed
atmospheric metallicities, respectively. We also suggest this
interpretation, along with extant observations of HMP stars, may provide
confirmation that the lower mass limit of the primordial initial stellar
mass function extends to below $1 M_\odot$.

In our current analysis, we could only compare the $\langle V_\phi
\rangle_{\rm med}$-[Fe/H] correlation for different types of stars, due
to the large extant errors in proper motion. However, the situation will
be greatly improved in the near future, when the Gaia satellite and
large ground-based telescopes provide truly enormous samples of disk and
halo stars with much more accurate kinematic and chemical information.
Such datasets will enable analyses of other chemo-dynamical correlations
-- e.g., correlations between the full velocity dispersion tensors and
surface metal abundances -- as well as more rigorous analyses of the
$\langle V_\phi \rangle_{\rm med}$-[Fe/H] relation.

\acknowledgments{
We thank the referee for critical but constructive comments on our manuscript. 
KH thanks Takafumi Sonoi, Masaomi Tanaka, Takuma Suda, and Akimasa
Kataoka for stimulating discussions. 
KH is supported by JSPS Research Fellowship for Young Scientists (23$\cdot$954). 
YY acknowledges partial support from the Grant-in-Aids of Scientific Research (17104002) 
of the Ministry of Education, Science, Culture and Sports of Japan. 
TCB acknowledges partial support for this work from grant PHY 08-22648: 
Physics Frontiers Center / JINA, awarded by the US National Science Foundation.  
YSL is a Tombaugh Fellow.}

\appendix

\section{Mock Catalogs of G/K-type dwarfs}

\subsection{Construction of Mock Catalogs}

In section \ref{mock}, we used a set of mock catalogs of G/K-type dwarfs
in order to investigate the reality of the observed offset in the $\langle V_\phi
\rangle_{\rm med}$-[Fe/H]. Here we present some
additional information on these mock catalogs. 

The observed (real) catalog of G/K-type dwarfs can be mathematically expressed as 
\begin{equation} \label{info real}
	\left\{ \left( T_{{\rm eff},i}, \log g_i, {\rm [Fe/H]}_i, d_i, \ell_i, b_i, v_{los, i}^{hel}, PM^{\ell}_{i}, PM^{b}_{i} \right) \; | \; i = 1, \cdots, N \right\}, \;\;{\rm (observed \; sample)}, 
\end{equation}
where $N$ is the total number of G/K-type dwarfs  
and the nine quantities for the $i$-th entry denote 
the effective temperature ($T_{{\rm eff},i}$), 
surface gravity ($\log g_i$), atmospheric metallicity (${\rm [Fe/H]}_i$), 
SSPP distance ($d_i$), Galactic longitude ($\ell_i$), Galactic latitude ($b_i$), 
heliocentric line-of-sight velocity ($v_{los, i}^{hel}$), 
proper motion in the $\ell$- and $b$-directions ($PM^{\ell}_{i}$, $PM^{b}_{i}$) 
of the $i$-th star. 
We generate each of our mock catalogs 
so that the $i$-th star in the mock catalog ($i = 1,\cdots, N$) 
has the same information on effective temperature, surface gravity, metallicity and 3-D position 
as those of the $i$-th star in our real (observed) catalog, 
while its 3-D velocity is assigned according to a given realistic
Galactic model, as well as to the assumed error models. 
Namely, the $j$-th mock catalog ($j = 1,\cdots, M$) can be expressed as 
\begin{equation} \label{info mock catalog}
	\left\{ \left( T_{{\rm eff},i}, \log g_i,  {\rm [Fe/H]}_i, d_i, \ell_i, b_i, v_{los, i, j}^{mock\mathchar`-obs, hel}, PM^{mock\mathchar`-obs, \ell}_{i, j}, PM^{mock\mathchar`-obs, b}_{i, j} \right) \; | \; i = 1, \cdots, N \right\}, 
	( j\mathchar`-{\rm th \; mock \; catalog)}. 
\end{equation}
In assigning the 3-D velocity information to the $i$-th star of a given mock catalog, we follow four steps:
\begin{itemize}
	\item {\bf Step 1} Assign the `true' [Fe/H] and the `true' 3-D position to the mock star;
	\item {\bf Step 2} Determine whether the $i$-th mock star belongs to halo or thick disk with a certain probability based on the `true' [Fe/H] and the `true' 3-D position;
	\item {\bf Step 3} Assign the `true' 3-D velocity according to the distribution function model of the halo and thick disk;
	\item {\bf Step 4} Decompose the `true' 3-D velocity into the line-of-sight velocity and proper motion and add realistic observational errors. 
\end{itemize}
Note that the `true' [Fe/H] and the `true' distance are only used as
internal variables, 
and do not appear explicitly in the mock catalogs. 
In the following, we describe each step in more detail.

In {\bf Step 1}, 
we assign the `true' metallicity 
\begin{equation}
	{\rm [Fe/H]}^{true}_{i, j} = {\rm [Fe/H]}_{i} + E_r ({\rm [Fe/H]}_i) 
\end{equation}
to the $i$-th star in the $j$-th mock catalog. 
Here, $E_r$ is the error-correcting term. 
Also, we assign the `true' distance 
\begin{equation}
	d^{true}_{i, j} = d_{i} \times \mathcal{N}(\mu_{dist}, 0.2)
\end{equation}
to the $i$-th star. 
Here, $\mathcal{N}(\mu, \sigma)$ is a random number generator that 
generate random numbers obeying a Gaussian distribution function 
with the mean of $\mu$ and the dispersion of $\sigma$. 
When there is no systematic error, we set $\mu_{dist} = 1$. 
The choice of $\sigma = 0.2$ is motivated by the 
$\sim 20$ \% random error in the distance estimation. 
We note that there is a tiny probability 
for the random number generator $\mathcal{N}$ to generate negative numbers. 
In that case, we set $d^{true}_{i, j} = 0.1\;{\rm kpc}$. 

In {\bf Step 2}, 
we randomly assign a flag of $H$ (halo) or $TD$ (thick disk) to the $i$-th star 
with a probability of $p$ or $(1-p)$, respectively, 
where $p$ is defined by 
\begin{equation} \label{def p}
p = \frac{f_{{H}}({\rm[Fe/H]}^{true}_{i,j})}{f_{H}({\rm[Fe/H]}^{true}_{i,j}) + f_{TD}({\rm[Fe/H]}^{true}_{i,j}) \times D(d^{true}_{i,j} \cdot \sin b_i)}. 
\end{equation}
Here, 
$D(z)$ describes the density profile of the thick disk as a function of the vertical distance from the Galactic disk plane $|z|$; and 
$f_{H}$ and $f_{TD}$ denote the MDFs at the Galactic plane ($z = 0 \; {\rm kpc}$) of halo and thick disk stars, respectively. 
We assume the following models for these functions: 
\begin{eqnarray}
f_{k}({\rm[Fe/H]}) &=& F_k \times \frac{1}{\sqrt{2 \pi} \sigma_k} \exp \left[ - \frac{1}{2} {\left( \frac{{\rm[Fe/H]} - \mu_k}{\sigma_k} \right)}^2\right], \;\; (k = H, TD), \\  
D(z) &=& \exp \left[- \frac{|z|}{h_{z}} \right], \;\; h_z = 1.0 \;{\rm kpc}, 
\end{eqnarray}
and we set $(F_H, \mu_H, \sigma_H)=(0.001, -1.5, 0.3)$, $(F_{TD}, \mu_{TD}, \sigma_{TD})=(0.040, -0.6, 0.2)$. 
We note here that we do not expect significant contribution from the thin disk, 
because of our metallicity cut of [Fe/H]$<-0.5$.

In {\bf Step 3}, 
we assign the `true' 3-D velocity of the $i$-th star in the usual Galactic cylindrical coordinate system by
\begin{equation} \label{input velocity}
	\left( V^{true}_{R, i, j}, V^{true}_{\phi, i, j}, V^{true}_{z, i, j} \right) = 
		\begin{cases}
			\left( \mathcal{N}(0, 80), \; \mathcal{N}(50, 150), \; \mathcal{N}(0, 70) \right) \;{\rm km\; s^{-1}}  & ( flag = H), \\
			\left( \mathcal{N}(0, 30), \; \mathcal{N}(180, 30), \; \mathcal{N}(0, 30) \right) \;{\rm km\; s^{-1}}  & ( flag = TD). 
	\end{cases}
\end{equation} 
Here, it is assumed that the mean rotational velocities of halo and thick disk 
are $50 \; {\rm km\;s^{-1}}$ and $180 \; {\rm km\;s^{-1}}$, respectively.  

In {\bf Step 4}, 
we calculate the `true' heliocentric line-of-sight velocity $v_{los, i, j}^{true, hel}$ 
and the `true' proper motion $PM^{true, \ell}_{i, j}$, $PM^{true, b}_{i, j}$ of the $i$-th mock star in $j$-th mock catalog, 
by using the `true' distance $d^{true}_{i, j}$ and 
by assuming the same LSR and peculiar solar velocity as those described in section \ref{Kinematical Information}. 
Then we calculate the `mock-observed' values for these quantities by using 
\begin{eqnarray}
v_{los, i, j}^{mock\mathchar`-obs, hel} &=& v_{los, i, j}^{true, hel}  + \mathcal{N}(0, 2) \;{\rm km\; s^{-1}}, \\
PM^{mock\mathchar`-obs, \ell}_{i, j} &=& PM^{true, \ell}_{i, j} + \mathcal{N}(0, 3.5) \;{\rm mas\; yr^{-1}}, \\
PM^{mock\mathchar`-obs, b}_{i, j} &=& PM^{true, b}_{i, j}  + \mathcal{N}(0, 3.5) \;{\rm mas\; yr^{-1}}.
\end{eqnarray}
We note that the adopted errors in the line-of-sight velocity ($2\;{\rm km\; s^{-1}}$) and 
in proper motion ($3.5\;{\rm mas\; yr^{-1}}$) are typical values in our real sample.

\begin{table}[htb]
  \begin{center}
  \caption{Metallicity error models in the mock catalogs}
  \begin{tabular}{|l||c|c|c|} \hline
  	Model    & $E_r ({\rm [Fe/H]}) $ for G-type dwarfs & $E_r ({\rm [Fe/H]}) $ for K-type dwarfs  \\ \hline \hline
	Model A & $\mathcal{N}(0, 0.2)$ & $\mathcal{N}(0, 0.2)$  \\
	Model B & $\mathcal{N}(0, 0.2)$ & $\mathcal{N}(0, 0.3)$  \\
	Model C & 
	$\begin{cases}
		\mathcal{N}(0, 0.2) & ({\rm [Fe/H]} \ge -1) \\
		\mathcal{N}(0, 0.2) + 0.2 \times ({\rm [Fe/H]} + 1.0) & ({\rm [Fe/H]} < -1) 
	\end{cases}$
	& 
	$\begin{cases}
		\mathcal{N}(0, 0.2) & ({\rm [Fe/H]} \ge -1) \\
		\mathcal{N}(0, 0.2) - 0.2 \times ({\rm [Fe/H]} + 1.0) & ({\rm [Fe/H]} < -1)
	\end{cases}$ \\ \hline
 \end{tabular}
 \label{error model}
 \end{center}
\end{table}

\subsection{Mock Observations of Mock Catalogs}

Observational errors in [Fe/H] and distance are the most important
factors that affect the derived $\langle V_\phi \rangle_{\rm
med}$-[Fe/H] relations. In order to evaluate how observational errors in
[Fe/H] affect our analyses, we consider three models of the error-correcting
term $E_r ({\rm [Fe/H]})$, as presented in Table \ref{error model}. Among
these models, Model A does not carry a spectral-type dependence in the
[Fe/H] error, while Models B and C are designed to carry this spectral-type
dependence. We also consider three types of systematic errors in
distance, by adopting $\mu_{dist}=0.8, 1.0,$ or $1.2$. For each of the
nine combinations of the error models (three for the [Fe/H] error and three for
the distance error), we generate 100 mock catalogs and derive the
$\langle V_\phi \rangle_{\rm med}$-[Fe/H] relations. 

\begin{figure*}
\begin{center}
	\includegraphics[angle=-90, width=0.3\columnwidth] {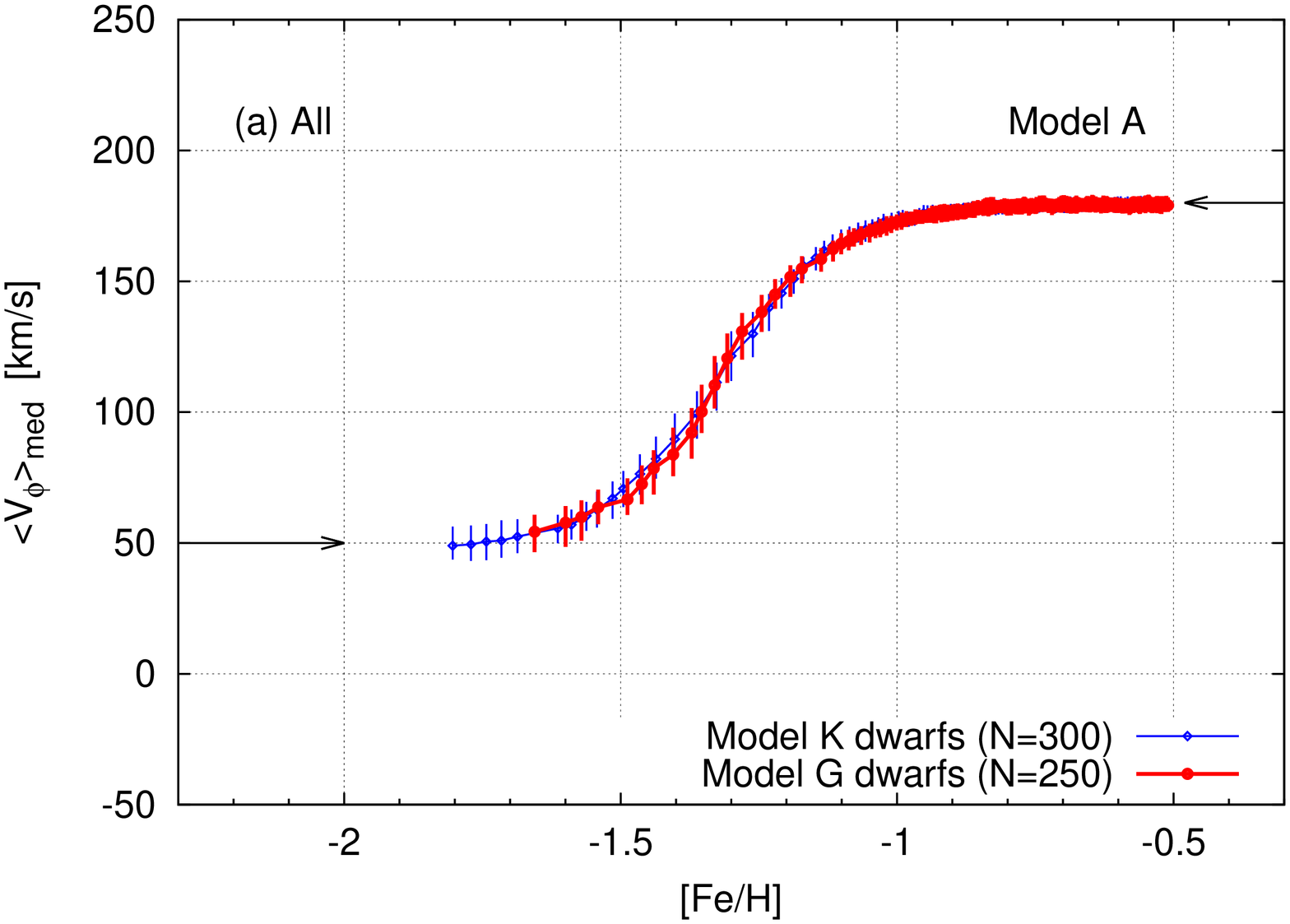}
	\includegraphics[angle=-90, width=0.3\columnwidth] {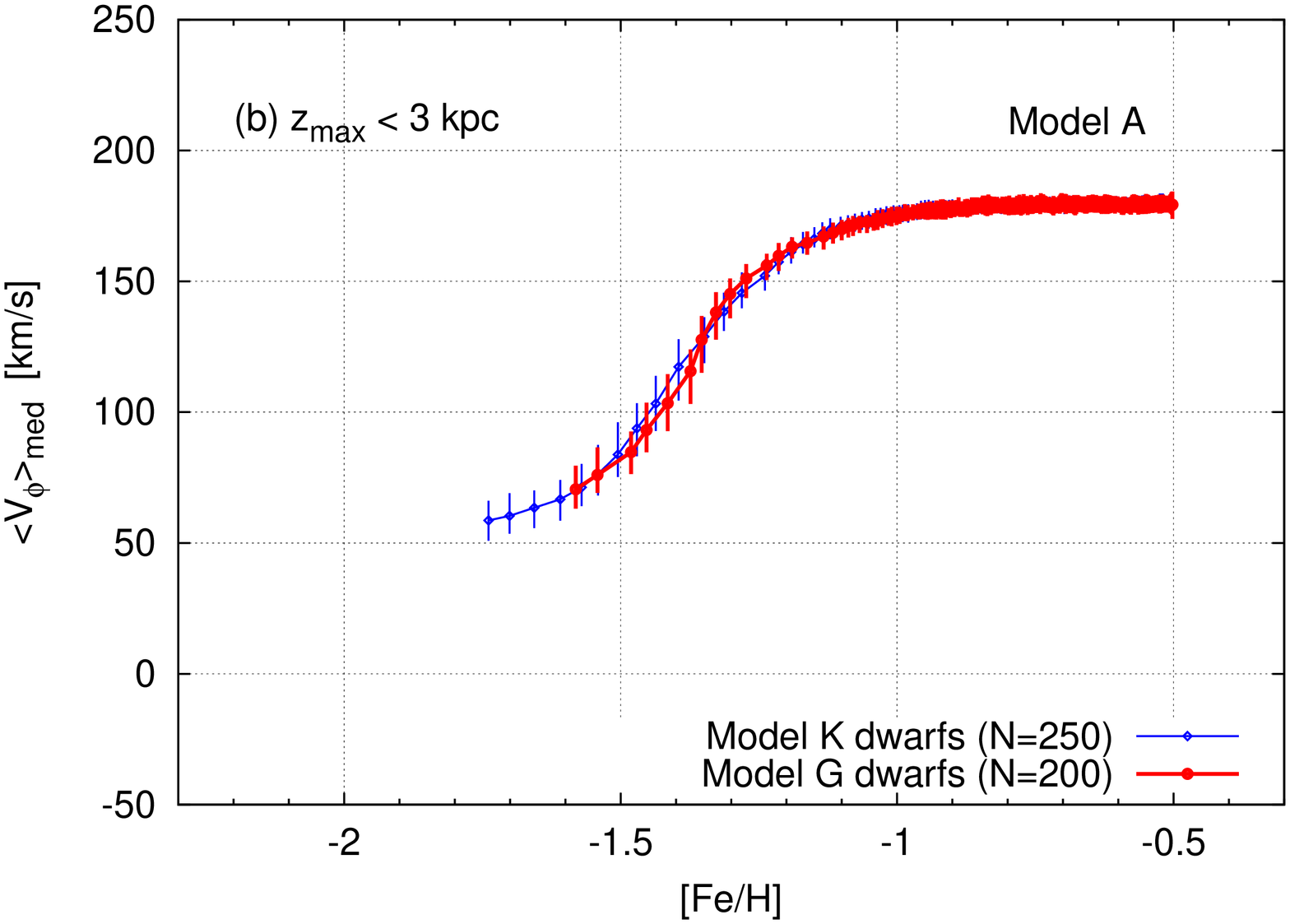}
	\includegraphics[angle=-90, width=0.3\columnwidth] {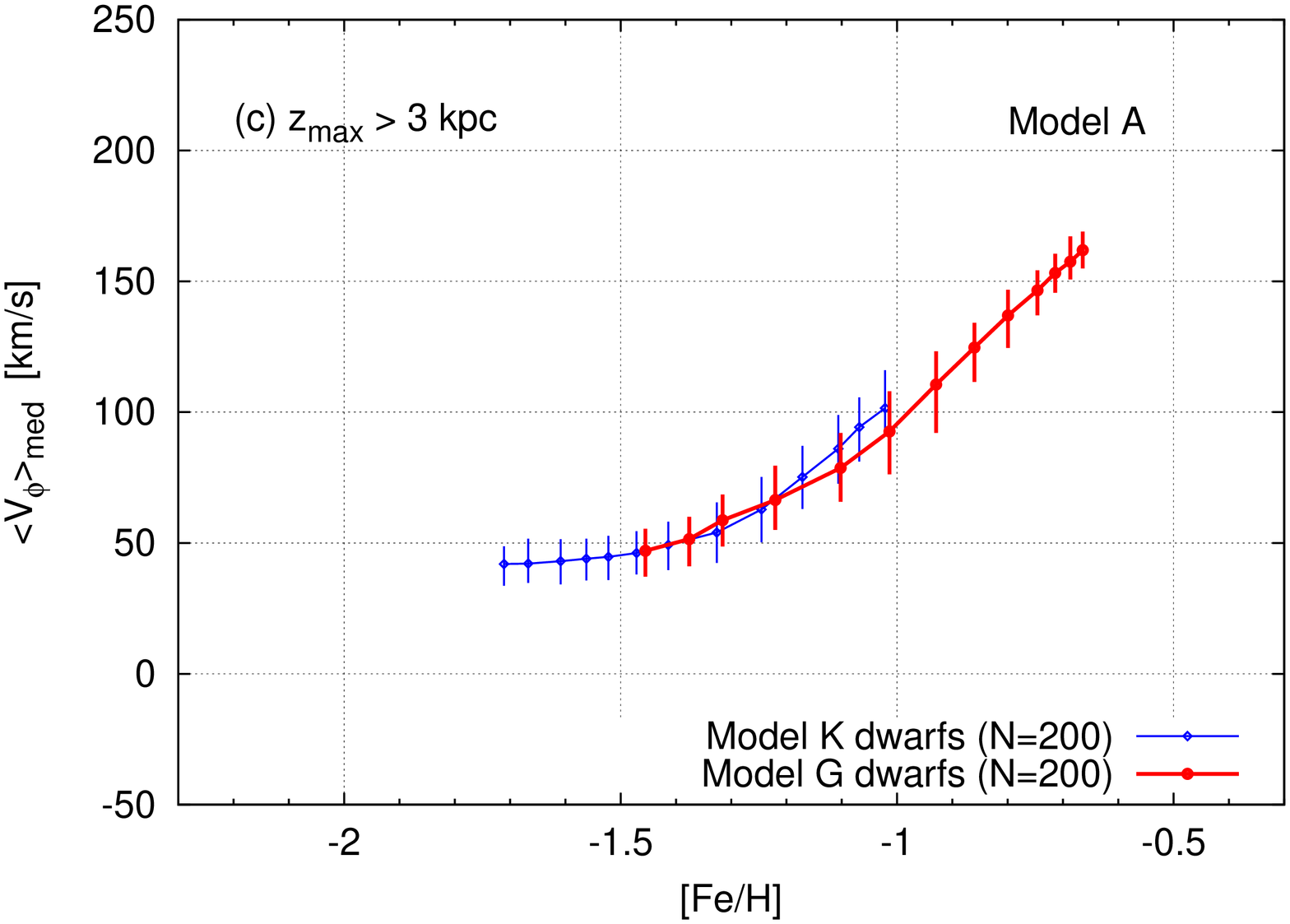}
\end{center}
\caption{
The $\langle V_\phi \rangle_{\rm med}$-[Fe/H] relations for 100 mock
catalogs with $\mu_{dist} = 1.0$. The implemented error in [Fe/H] is
described by Model A for these mock catalogs. The red and blue lines
indicate the median curve of 100 $\langle V_\phi \rangle_{\rm
med}$-[Fe/H] relations for G- and K-type dwarfs, respectively. (a) The
result for the full sample without $z_{\rm max}$ cut applied, which
corresponds to Figure \ref{fig 1 pm}(c). The horizontal arrows at $50\;
{\rm km\;s^{-1}}$ and $180\;{\rm km\;s^{-1}}$ indicate the mean
velocities of halo and thick-disk stars, respectively, expected from the
input distribution function model [see equation (\ref{input velocity})].
(b) The result for the subsample of stars with $z_{\rm max} < 3\;{\rm
kpc}$, which corresponds to Figure \ref{fig 2 zmax}(a). (c) The result
for the subsample of stars with $z_{\rm max} > 3\;{\rm kpc}$, which
corresponds to Figure \ref{fig 2 zmax}(b). }
\label{fig model A}
\end{figure*}

\begin{figure*}
\begin{center}
	\includegraphics[angle=-90, width=0.3\columnwidth] {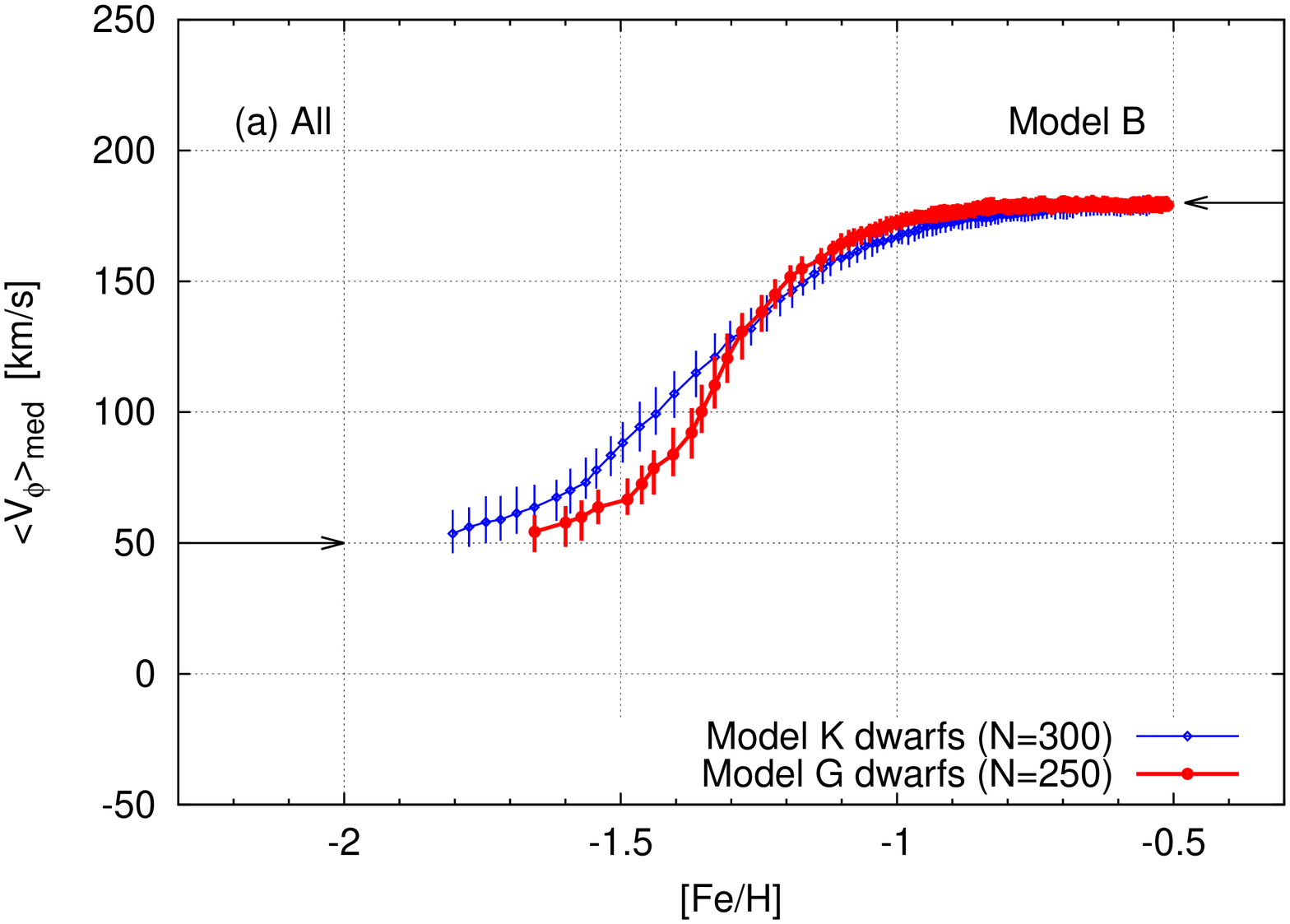}
	\includegraphics[angle=-90, width=0.3\columnwidth] {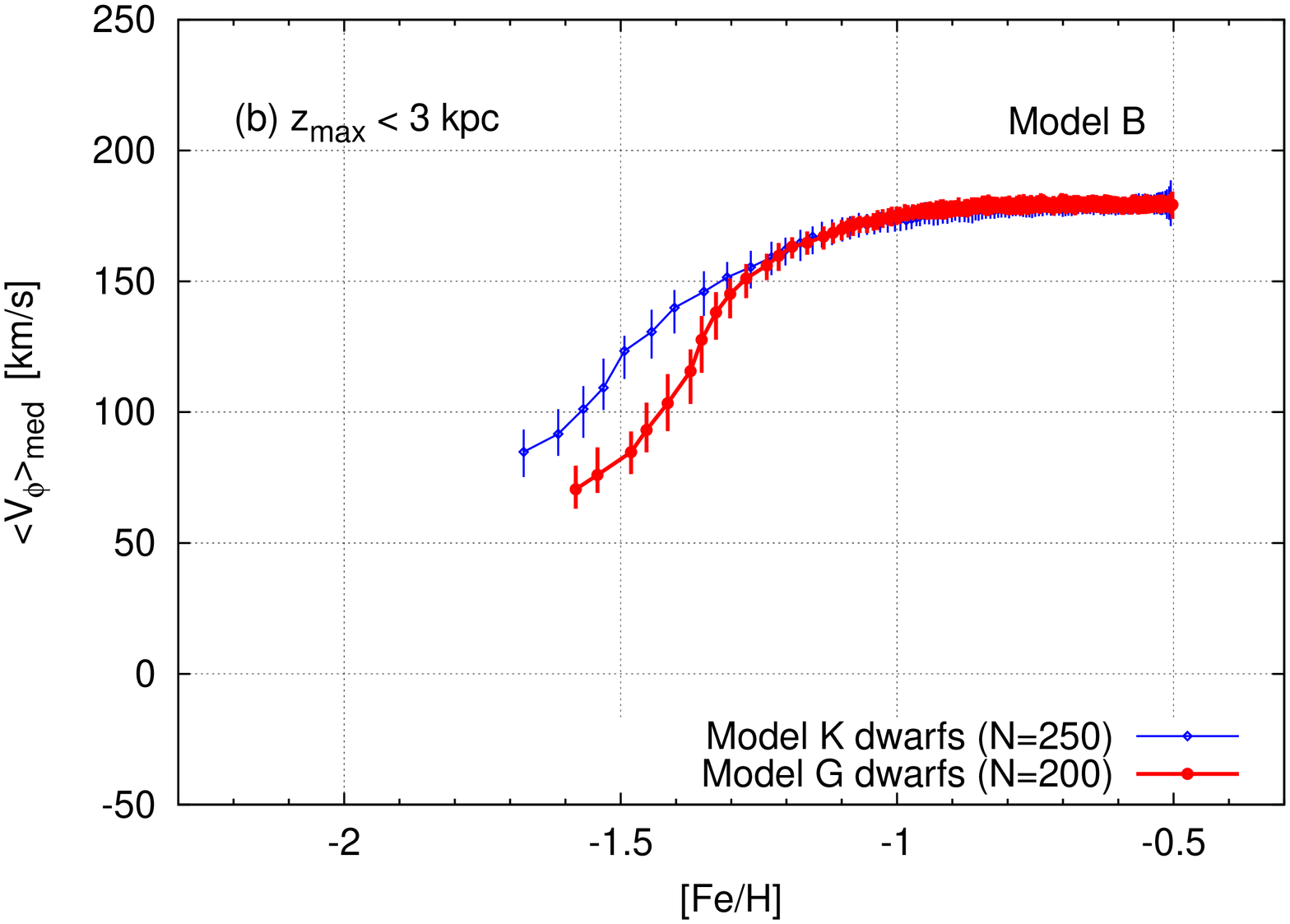}
	\includegraphics[angle=-90, width=0.3\columnwidth] {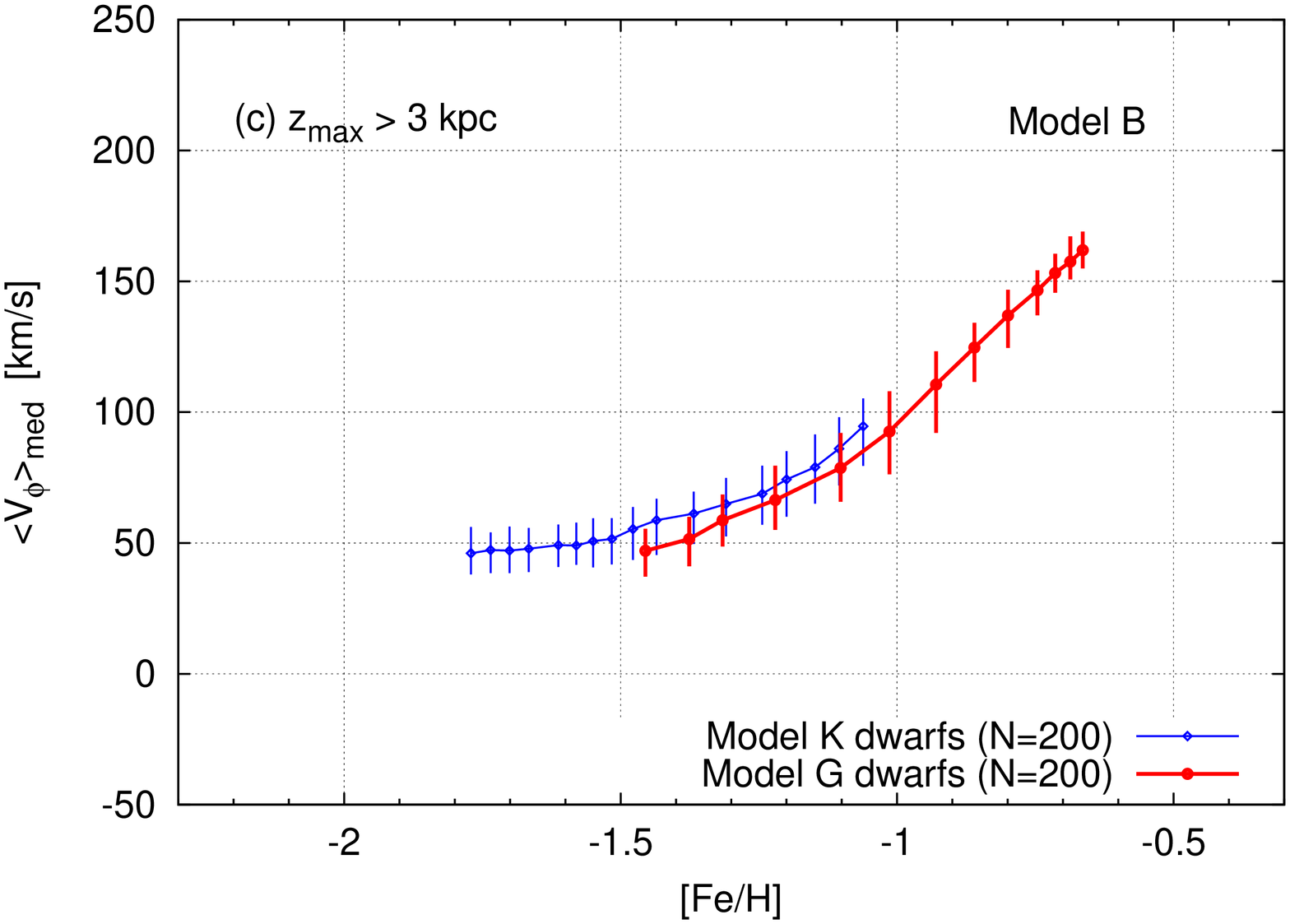}
\end{center}
\caption{
The same as in Figure \ref{fig model A}, but for 100 mock catalogs in which the error in [Fe/H] is described by Model B. 
}
\label{fig model B}
\end{figure*}

\begin{figure*}
\begin{center}
	\includegraphics[angle=-90, width=0.3\columnwidth] {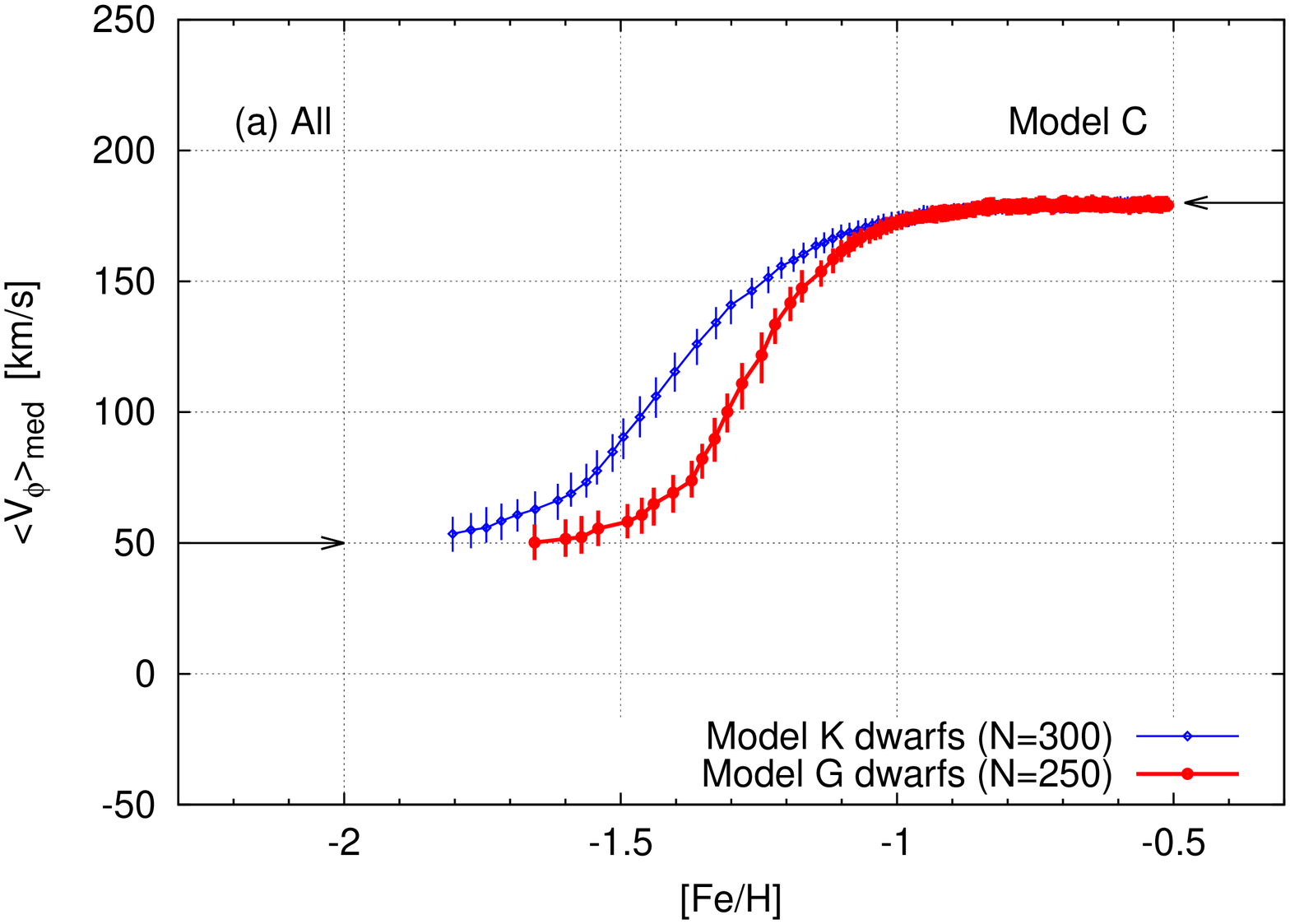}
	\includegraphics[angle=-90, width=0.3\columnwidth] {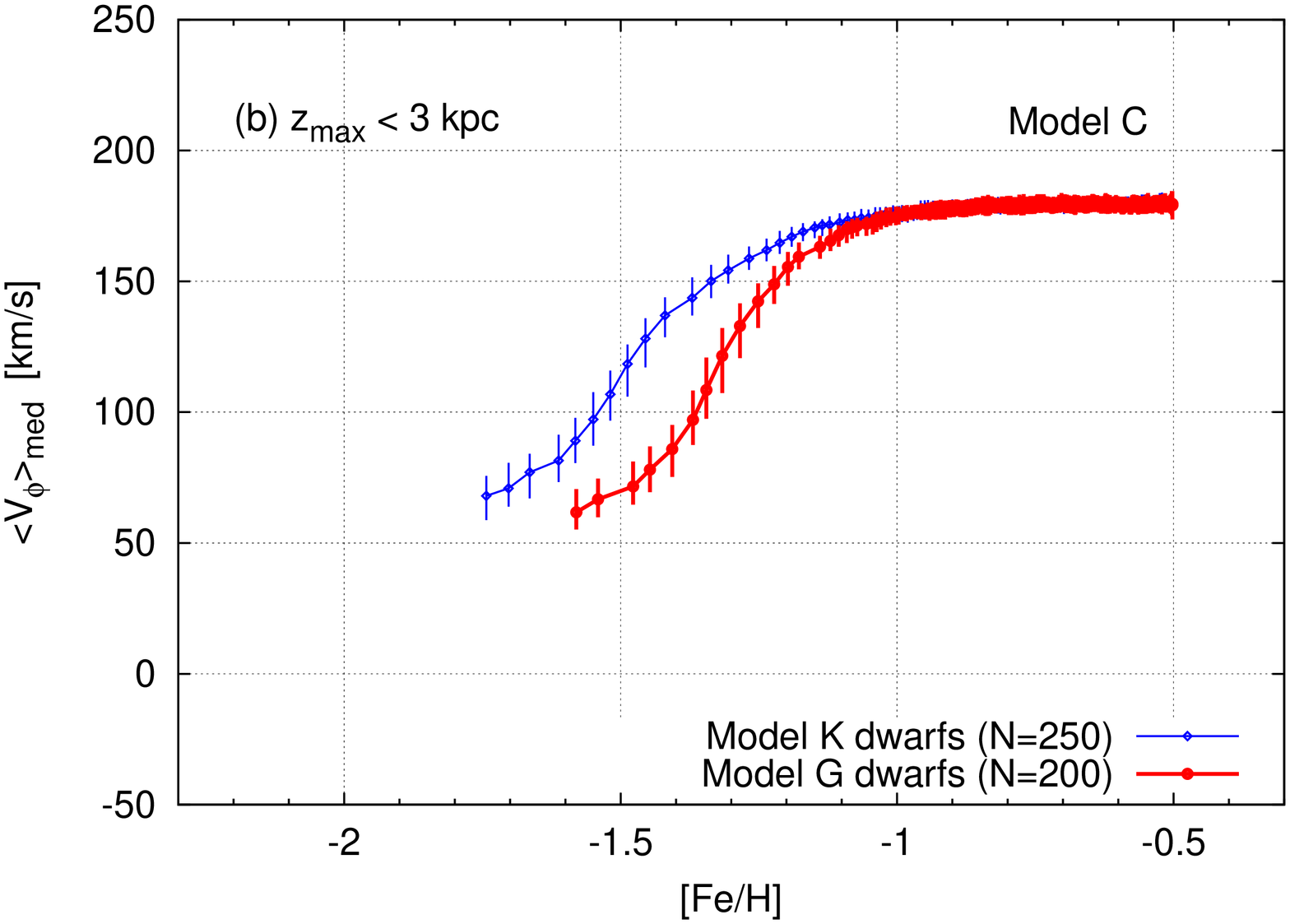}
	\includegraphics[angle=-90, width=0.3\columnwidth] {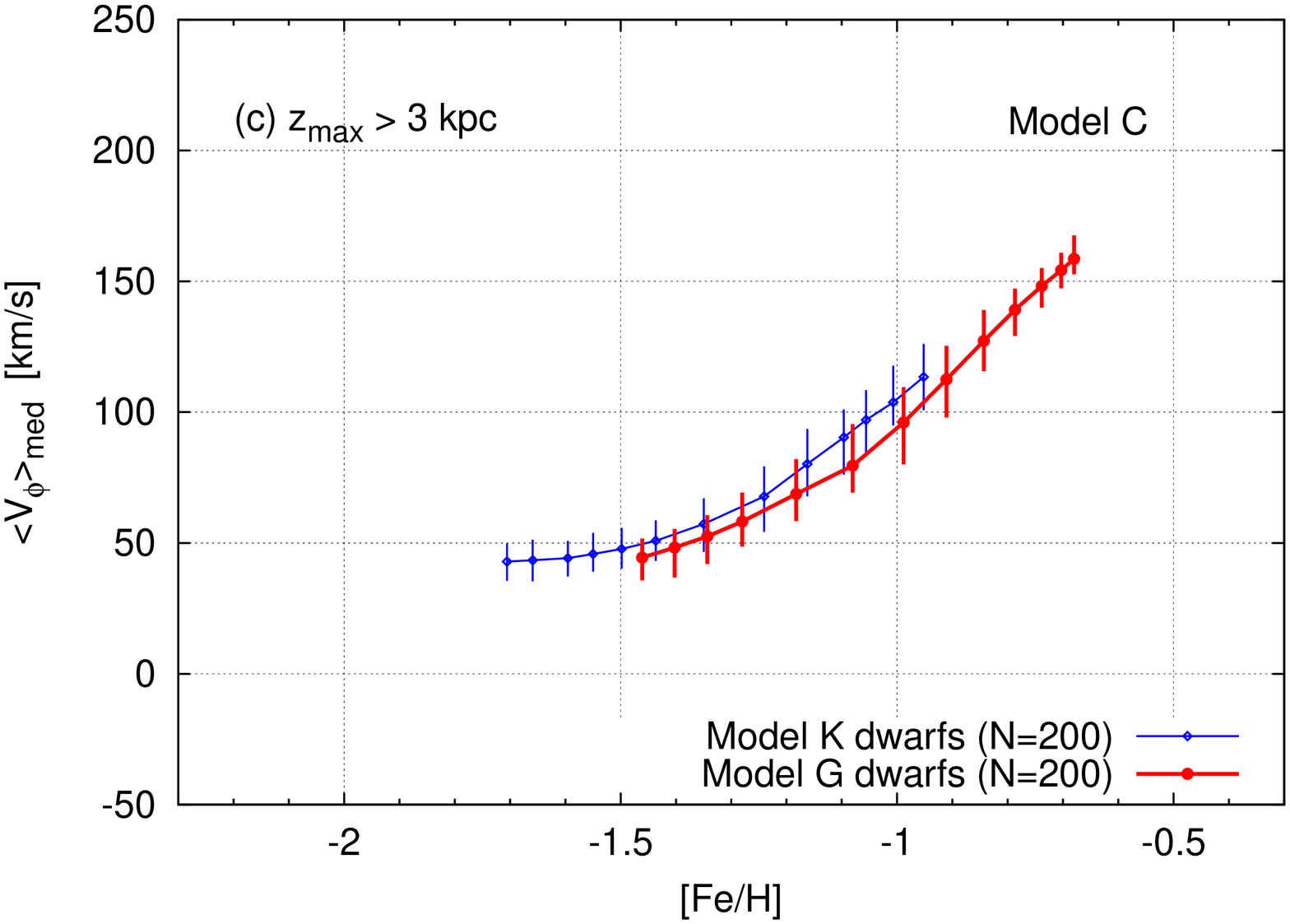}
\end{center}
\caption{
The same as in Figure \ref{fig model A}, but for 100 mock catalogs in which the error in [Fe/H] is described by Model C.}
\label{fig model C}
\end{figure*}

In Figures \ref{fig model A}, \ref{fig model B}, and \ref{fig model C},
we show the results for Models A, B, and C, respectively, with
$\mu_{dist}$ fixed to be $1.0$. In these figures, we show the median
curve of 100 $\langle V_\phi \rangle_{\rm med}$-[Fe/H] relations derived
from the mock catalogs. The panels (a), (b), and (c) correspond to
Figures \ref{fig 1 pm}(c), \ref{fig 2 zmax}(a), and \ref{fig 2 zmax}(b),
respectively. 

Figure \ref{fig model A} suggests that, if the observational errors in
[Fe/H] do not possess a spectral-type dependence (Model A), the
resultant $\langle V_\phi \rangle_{\rm med}$-[Fe/H] relations are
statistically identical for G- and K-type dwarfs. On the other hand,
Figures \ref{fig model B} and \ref{fig model C} suggest that we expect
an offset in the $\langle V_\phi \rangle_{\rm med}$-[Fe/H] relation for
the full sample and the low-$z_{\rm max}$ subsample, if there is a
spectral-type dependence in the observational errors in [Fe/H] (Model B
or C), although the offset is not very clear for the high-$z_{\rm max}$
subsample (due to the small sample size).

By comparing the panels (a) and (b) in Figures \ref{fig model A},
\ref{fig model B}, and \ref{fig model C}, it is suggested that, whenever
we see an offset in the full sample [panel (a)], we should also see a similar offset
in the low-$z_{\rm max}$ subsample [panel (b)]. In other words, if
the observed offset in Figure \ref{fig 1 pm}(c) is due to the
observational errors in [Fe/H], then a similar offset is also expected
in Figure \ref{fig 2 zmax}(a) as well. Thus, it seems that the
detected offset in Figures \ref{fig 1 pm}(c) and \ref{fig 2 zmax}(b), as
well as the non-existence of the offset in Figure \ref{fig 2 zmax}(a),
are due to neither an observational error in [Fe/H] nor its
spectral-type dependence. 

\begin{figure}[!htbp]
  \begin{center}
    \leavevmode
	\includegraphics[angle=-90,width=0.3\columnwidth]{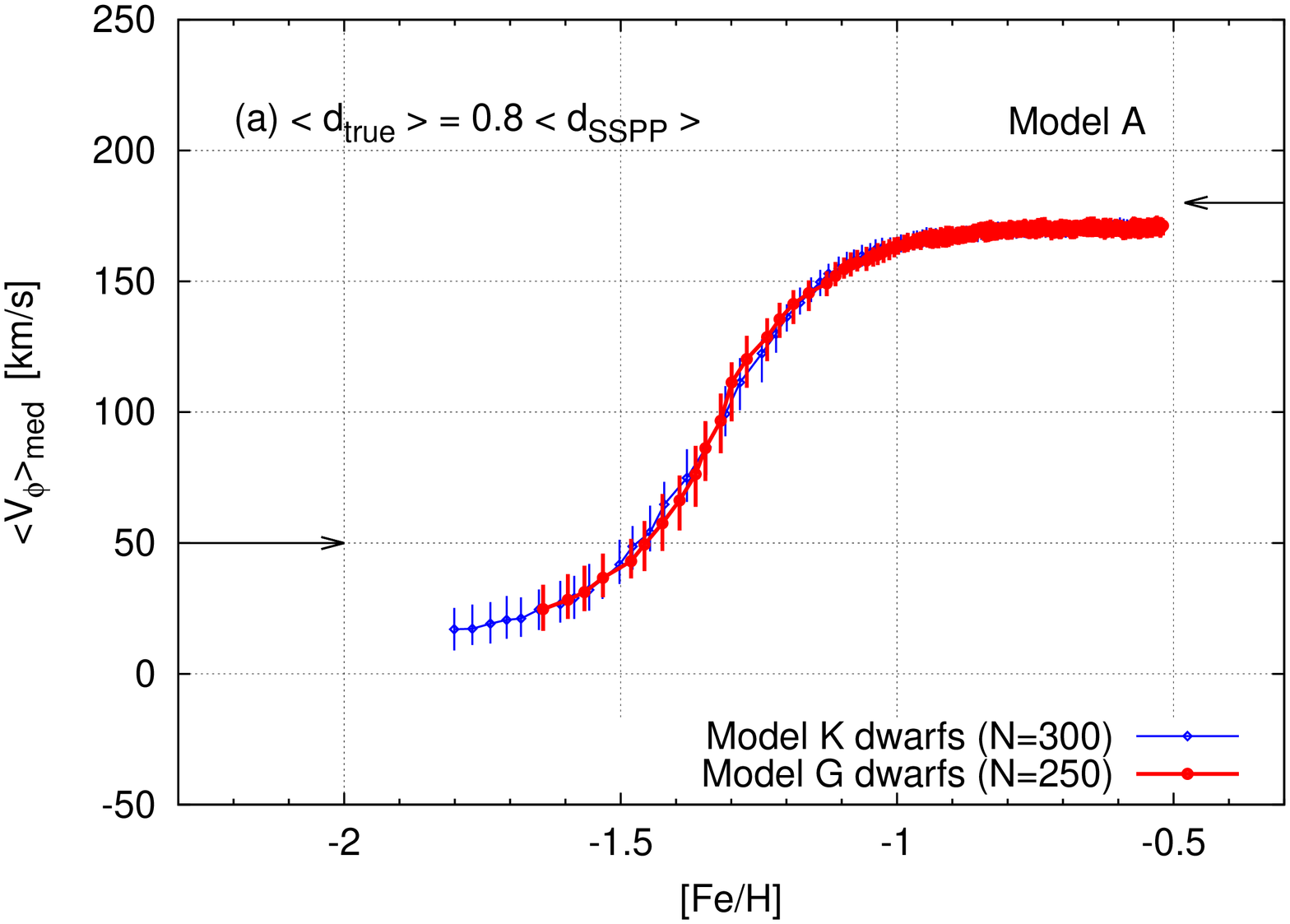} 
	\includegraphics[angle=-90,width=0.3\columnwidth]{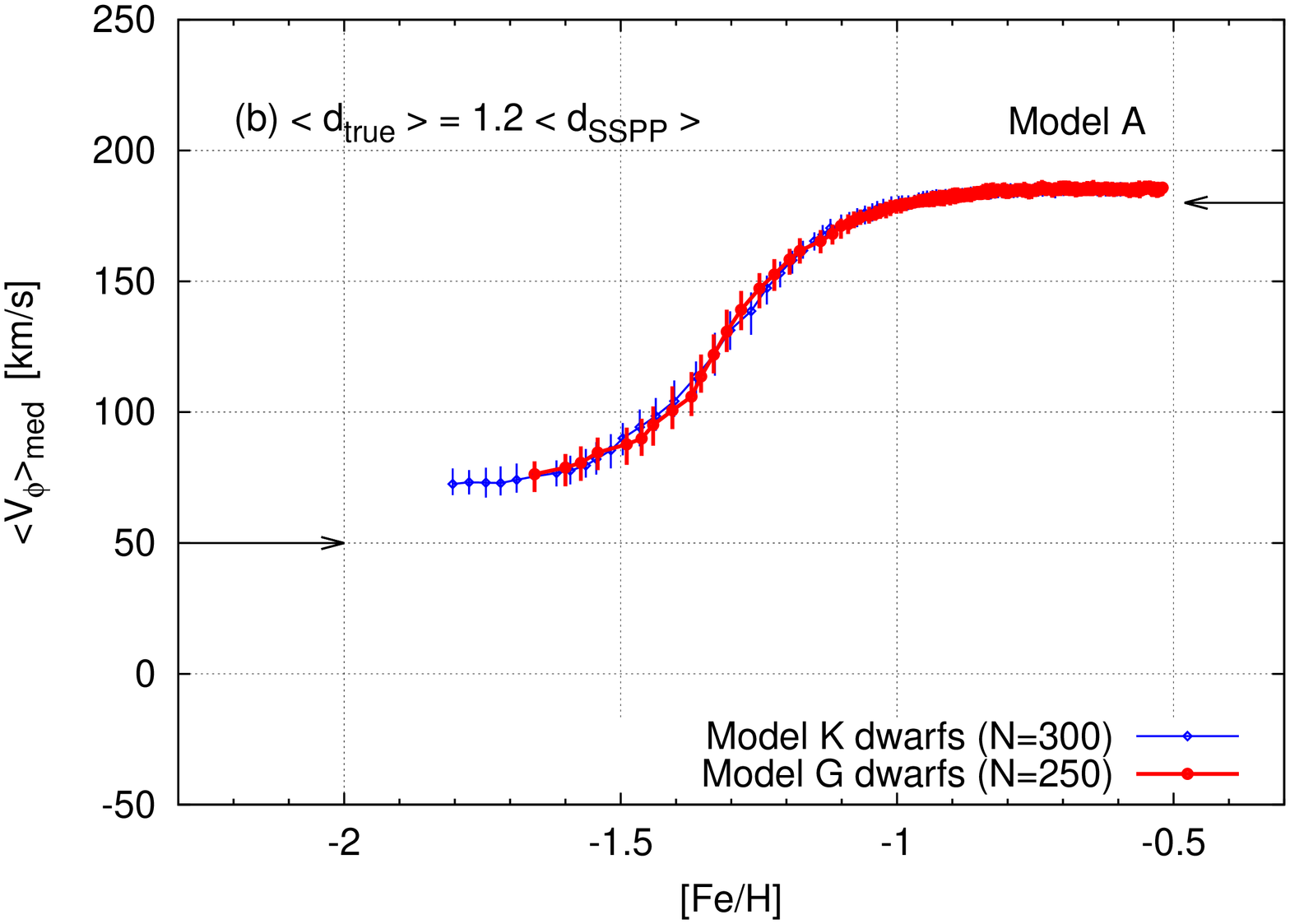} 
    \caption{
    The same as in Figure \ref{fig model A}(a), 
    but for $\mu_{dist}= 0.8$ (left-hand panel) and $\mu_{dist}= 1.2$ 
    (right-hand panel). 
    }
    \label{fig dist}
  \end{center}
\end{figure}

If we further vary the value of $\mu_{dist}$ for G- and/or K-type
dwarfs, the results are essentially the same as in the case of
$\mu_{dist}=1.0$. The only difference is in the overall location of the
$\langle V_\phi \rangle_{\rm med}$-[Fe/H] curves. If $\mu_{dist}$ is set
to be $0.8$ and the SSPP distance is overestimated, the resultant curve
of the $\langle V_\phi \rangle_{\rm med}$-[Fe/H] relation is shifted
downwards (toward lower $\langle V_\phi \rangle_{\rm med}$), due to the
overestimate of the relative velocities of the sample stars with respect
to the Sun. On the other hand, if $\mu_{dist}$ is set to be $1.2$ and
the SSPP distance is underestimated, the resultant curve of $\langle
V_\phi \rangle_{\rm med}$-[Fe/H] relation is shifted upwards. The effect
of $\mu_{dist}$ is most prominently seen at the low-metallicity tail of
the $\langle V_\phi \rangle_{\rm med}$-[Fe/H] relation. For example, if
we adopt Model A, the plateau value of $\langle V_\phi \rangle_{\rm
med}$ in the low-metallicity tail becomes $\simeq 20 \;{\rm km\;s^{-1}}$
when $\mu_{dist} = 0.8$, while this value becomes $\simeq 70 \;{\rm km\;
s^{-1}}$ when $\mu_{dist} = 1.2$ (see Figure \ref{fig dist}). The fact
that both the G- and K-type dwarfs exhibit $\langle V_\phi \rangle_{\rm
med} \simeq 50 \;{\rm km \; s^{-1}}$ at the low-metallicity tail in
Figure \ref{fig 1 pm}(c) suggests that there is no spectral-type
dependence in $\mu_{dist}$ in our sample. Therefore, we conclude that
the observed offset in the $\langle V_\phi \rangle_{\rm med}$-[Fe/H]
relation [Figures \ref{fig 1 pm}(c) and \ref{fig 2 zmax}(b)], or the
non-existence of it [Figure \ref{fig 2 zmax}(a)], are not due to
observational errors in [Fe/H] or distance.

\end{document}